\documentclass[10pt,a4paper,reqno]{amsproc}
\usepackage{longtable,cite}
\usepackage[a4paper, top=1in, bottom=1in, left=0.8in, right=0.8in]{geometry}
\usepackage{lscape}
\usepackage{multirow}
\usepackage[pdftex,bookmarks=true]{hyperref}
\newtheorem{theorem}{Theorem}[section]

\theoremstyle{definition}
\newtheorem{definition}[theorem]{Definition}

\theoremstyle{remark}

\numberwithin{equation}{section}
\usepackage{hyperref}
\usepackage{tikz}
\usepackage{array}
\usetikzlibrary{shapes, arrows}
\usetikzlibrary{chains,arrows.meta}
\usepackage{placeins}
\begin{document}
	\title [Analysis of Delay Differential Equations Using the Offset Linear Canonical Transform
	]{Analysis of Delay Differential Equations Using the Offset Linear Canonical Transform} 
	\author{Gita Rani Mahato}
	\author{Manab Kundu}
	
	\address{Department of Mathematics, SRM University AP, Amaravati-522240, India}
	
	\email{\hfill \break
		gitamahato1158$@$gmail.com (Gita Rani Mahato),
		\hfill \break manabiitism17$@$gmail.com (Manab Kundu-Corresponding author))}
	
	\thanks{Corresponding author: Manab Kundu}
	
	\date{}
    \keywords{ Fourier transform; Offset linear canonical transform; Delay differential equation ; }
		\maketitle
        
		\begin{abstract}
		 Motivated by the work of Ohira \cite{ohira1,ohira ft} and the advantages of the offset linear canonical transform (OLCT) over the Fourier transform (FT), this paper proposes an OLCT based framework for solving a class of delay differential equations. By exploiting the operational properties of the OLCT, the original delay differential equation is transformed into a Volterra-type delay integral equation in the transform domain and solved numerically using Brunner's method of steps \cite{brunner}. An explicit analytical solution is derived for a special case to investigate the effect of the delay parameter. A unified transform-domain formulation is also established by relating the proposed OLCT approach to its Fourier transform counterpart. Numerical and graphical results demonstrate the accuracy and effectiveness of the proposed method and validate it through comparisons with the Fourier transform based formulation of Ohira et al. \cite{ohira ft}. The proposed framework may provide a promising foundation for solving non linear delay differential equations involving transcendental terms, as both the OLCT and the resulting Volterra integral equation possess the essential properties required to handle non linearities and transcendental terms.
		\end{abstract}
	\section{Introduction}
 The past often influences the present evolution of a system. Such memory effects are common in biology, economics, engineering, and many other disciplines. To describe these phenomena, delay differential equations (DDEs) are used. Delay differential equations, also known as difference-differential equations, have their origins in the eighteenth century through the works of Laplace and Condorcet.\cite{Richard}. Since then, various types of DDEs have been developed to model different real-world processes. For example, retarded delay differential equations (RDDEs) contain delays only in the state variable, while neutral delay differential equations (NDDEs) contain delays in both the state variable and its derivative In particular, DDEs are employed in studying the spread and dynamics of infectious diseases \cite{neutral dde, Ruan}. 
\\
\\
Unlike ordinary differential equations, the behavior of a delay differential equation depends not only on its current state but also on its past states. As a result, the dynamics of the system become more complex. The presence of a delay can significantly affect the stability of the solution and may also produce oscillatory behavior that does not occur in systems without delays \cite{Menon,Gopalsamy1992}. The analysis of DDEs is often more challenging than that of ordinary differential equations due to the presence of delayed terms. As a result, various analytical and numerical methods have been developed to investigate their solutions and dynamical properties\cite{bellen, ai}. The Laplace transform provides an effective approach for solving linear DDEs. For instance, Yi et al. developed a Laplace transform approach for solving linear systems of DDEs using the matrix Lambert function method\cite{yi}. Gilbert Kerr et al. proposed a new analytical technique for obtaining exact solutions of systems of linear retarded and neutral delay differential equations \cite{lt fs, kerr}. Furthermore, Sherman et al. investigated several neutral and non-neutral DDEs to compare the symbolic computation capabilities of Maple and MATLAB\cite{sherman, sherman kerr}.In the Laplace-transform approach, obtaining explicit solutions often depends on the ability to compute the inverse transform in a suitable closed form, which may become difficult for complicated delay systems. Moreover, the delay terms frequently lead to transcendental characteristic equations, making direct analytical treatment challenging. Recently, Ohira et al. employed Fourier analysis to investigate the dynamical behavior of a delay differential equation \cite{ohira1}. As a continuation of this work, Ohira et al. employed the Fourier transform to obtain explicit solutions of a delay differential equation with a linear time-dependent coefficient. The solution was derived in the Fourier domain and reconstructed through the inverse Fourier transform, providing a new transform-based approach for solving delay differential equations \cite{ohira ft}. As Fourier transform (FT) has been generalized into a wide class of integral transforms with additional degrees of freedom. It is worthwhile to investigate whether these generalized transforms, through their enhanced flexibility together with their generalized operational properties, can provide an effective framework for the analysis and solution of delay differential equations constitutes an important research direction.
\\
\\ 
 Among the generalized integral transforms developed in recent years, the  offset linear canonical transform (OLCT) has attracted considerable attention due to its rich mathematical structure and broad applicability \cite{eigen, abe}. The OLCT is a six-parameter integral transform that generalizes several important transforms, including the Fourier transform, the fractional Fourier transform (FrFT), and the linear canonical transform (LCT), as special cases\cite{d wei}. The additional parameters provide greater flexibility in representing signals and functions in the transform domain, making the OLCT a promising transform \cite{stern}. The fact that FT, FrFT and LCT are special cases of the Offset Linear Canonical Transform (OLCT) naturally motivates to investigate whether the additional degrees of freedom, enhanced flexibility in handling time shifts, frequency shifts, scaling, and chirp modulation, together with the generalized operational properties of the OLCT, provide a more effective framework for the analysis and solution of delay differential equations.
\\ \\
Despite of numerous advantages and significant progress in theoretical development and applications, the potential of the OLCT for the analysis and solution of delay differential equations remains unexplored. To the best of the authors' knowledge, an OLCT based framework for solving delay differential equations and establishing connections with Volterra-type delay integral equations has not been investigated in the existing literature. In particular, while Fourier and Laplace transform based techniques have been successfully employed for various classes of DDEs, an OLCT based framework capable of incorporating these approaches within a unified transform setting has not been adequately investigated. Since delayed systems exhibit complex memory-dependent dynamics, the availability of a more general transform framework may provide additional analytical flexibility and new perspectives for studying their solution behavior.
\\ \\
In the direction of the analysis and solution of delay differential equations (DDEs), our broader objective is to develop an effective framework for nonlinear DDEs involving transcendental terms. The Offset Linear Canonical Transform (OLCT), owing to its powerful convolution and correlation properties \cite{wei}, provides a promising mathematical framework for tackling such problems. Moreover, the application of the OLCT transforms the original DDE into an equivalent Volterra integral equation, which offers a natural setting for handling transcendental nonlinearities \cite{hermann}. As an initial step toward this broader objective, the present work focuses on linear delay differential equations. The effectiveness and applicability of the proposed approach are demonstrated through comparisons with existing results, while a detailed error analysis is performed to establish its accuracy and reliability.
\\ \\
In the present work, we develop an OLCT-based methodology for a class of delay differential equations. By exploiting the operational properties of the OLCT, the original delay differential equation is transformed into a Volterra-type delay integral equation in the transform domain. The Volterra integral formulation gives an equivalent form of the original problem, simplifies the analysis, and provides a convenient framework for developing stable numerical solutions\cite{burton}. This reformulation establishes a connection between delay differential equations, delay integral equations, and generalized transform theory. Furthermore, the relationship between the OLCT and the Fourier transform is utilized to derive alternative solution representations, thereby demonstrating how classical Fourier-based formulations arise as special cases of the proposed framework.
\\ \\
The main contributions of this paper are summarized as follows:
\begin{itemize}
\item Develop an OLCT-based framework for the analysis and solution of a class of delay differential equations.

\item  The original delay differential equation  transforms into a Volterra-type delay integral equation in the OLCT domain and construct its solution using Brunner's method of steps\cite{brunner}.

\item Derive explicit analytical solutions for a special case of the proposed model and investigate the influence of the delay parameter on the solution behavior.

\item  Establish a connection between the OLCT formulation and the corresponding Fourier-transform-based representation, thereby providing a unified transform-domain perspective.

\item The present numerical simulations and graphical illustrations that demonstrate the effectiveness of the proposed approach and highlight the impact of delay and transform parameters on the resulting solutions.

 \item Also validate the proposed OLCT framework through graphical and numerical comparisons between the direct OLCT formulation and the Fourier-transform-based OLCT formulation.

 \item The present graphical and numerical comparisons between the proposed OLCT solution and the Fourier-transform-based solution of Ohira \textit{et al.}\cite{ohira ft}, demonstrating the consistency and effectiveness of the proposed framework.

\end{itemize}
The organization of this paper is as follows. Section 2 presents the preliminary definitions and properties. Section 3 contains the main results. Section 4 presents a comparison between the proposed OLCT solution and the Fourier-transform-based solution. Section 5 provides the error analysis together with graphical and numerical comparisons. Finally, Section 6 contains the discussion and conclusion.

	\section{Preliminaries}
In this section, we discuss the fundamental definitions and properties of the Fourier transform, the OLCT, and the definition of delay differential equations.

	\begin{definition} \cite{ohira ft}
		Let the function $f \in L^1(\mathbb{R})$. Then Fourier transform of $f$  is  defined as 
		\begin{eqnarray}
			({F}f)(u) = \frac{1}{\sqrt{2\pi}} \int_{\mathbb{R}} e^{-itu}f(t)dt,\hspace{3mm} \forall u\in \mathbb{R}.
		\end{eqnarray}
		The inverse Fourier transform can be written as
		\begin{eqnarray}
			f(t)=\frac{1}{\sqrt{2\pi}} \int_{\mathbb{R}} e^{itu}({F}f)(u)du,\hspace{3mm} \forall t\in \mathbb{R}.
		\end{eqnarray}
	\end{definition}
	\begin{definition} \cite{eigen} Let $f\in   L^1(\mathbb{R}) $. Then OLCT of $f$ can be defined as  
		\begin{eqnarray}\label{2.3}
			\small
			\mathcal{O}_M[f(t)](u)=
			\begin{cases}
				\int_{\mathbb{R}}f(t)h_M(t,u)dt & \hspace{-4mm}, ~~ b\neq0\\
				\sqrt{d}~e^{i\frac{cd}{2}(u-u_0)^2+i\omega_0}f[d(u-u_0)] & \hspace{-0.3cm}, ~~b=0, 
			\end{cases}
		\end{eqnarray}
		where
		\begin{eqnarray}
			h_M (t ,u)= \sqrt{\frac{1}{2\pi i b}}~e^{\frac{i}{2b}(a t^2 + 2 t (u_0-u)-2u(du_0-b\omega_0)+ du^2+ du_0^2)},
		\end{eqnarray}
		is the kernel of OLCT, with	$ M= a, b, c, d, u_0, \omega_0 \in \mathbb{R}$ and $ad - bc = 1$. Here, we will restrict our focus to the OLCT case where 
		$b\neq0$ .
		Then its inverse transformation can be defined as 
		\begin{eqnarray}
			f(t)=  \Big(\mathcal{O}_{M^{-1}} (\mathcal{O}_{M} f)\Big)(t)= T \int_{\mathbb{R}}  \mathcal{K}_{M^{-1}} (t, u) (\mathcal{O}_{M} f)(u) du,
		\end{eqnarray}
		where 
		\begin{eqnarray*}
			&&{M^{-1}=(d,-b,-c,a,b\omega_0-du_0,cu_0-a\omega_0) }
			~\text{and} \\ && T= e^{\frac{i}{b}(cdu_0^2-2adu_0\omega_0+ab\omega_0^2)}.
		\end{eqnarray*}
		
	\end{definition}

	\subsection{Relation between FT and OLCT}\label{relation between FT and OLCT}
	\begin{eqnarray*}
		\mathcal{O}_M[f(t)](u)&=&\frac{1}{\sqrt{2\pi i b}}\int_{\mathbb{R}}e^{\frac{i}{2b}(a t^2 + 2 t (u_0-u)-2u(du_0-b\omega_0)+ du^2+ du_0^2)}f(t)\ dt\\
		&=&\frac{1}{\sqrt{2\pi i b}}e^{\frac{i}{2b}(-2u(du_0-b\omega_0)+ du^2+ du_0^2)}\int_{\mathbb{R}}e^{\frac{i}{2b}(a t^2 + 2 t (u_0-u))}f(t)\ dt
		\\&=&\frac{1}{\sqrt{2\pi i b}}e^{\frac{i}{2b}(-2u(du_0-b\omega_0)+ du^2+ du_0^2)}\int_{\mathbb{R}}e^{\frac{-i}{b}tu} f(t)e^{\frac{i}{2b}(a t^2 + 2 t u_0)}\ dt\\&=&\frac{1}{\sqrt{i b}}e^{\frac{i}{2b}(-2u(du_0-b\omega_0)+ du^2+ du_0^2)}\mathcal{F}( f(t)e^{\frac{i}{2b}(a t^2 + 2 t u_0)})(\frac{u}{b})\\&=& C e^{E(u)}\mathcal{F}[\hat f](\frac{u}{b})
			\end{eqnarray*}
			where $e^{E(u)}=e^{\frac{i}{2b}(-2u(du_0-b\omega_0)+ du^2+ du_0^2)}$,  $ C= \frac{1}{\sqrt{i b}}$ and $\hat{f}(t)=f(t)e^{\frac{i}{2b}(a t^2 + 2 t u_0)}$ 
			\subsection{Properties of FT}
	\begin{itemize}
	\item \textbf{Linearity Property:}
	\begin{eqnarray}
		\mathcal{F}\{a f(t) + b g(t)\}(u)
		&=&
		a\,\mathcal{F}\{f(t)\}(u) + b\,\mathcal{F}\{g(t)\}(u)
	\end{eqnarray}

	\item \textbf{Shifting Property (Time Shift):}
	\begin{eqnarray}
		\mathcal{F}\{f(t-\tau)\}(u)
		&=&
		e^{-i u \tau}\,\mathcal{F}\{f(t)\}(u)
	\end{eqnarray}

	\item \textbf{Differentiation Property:}
	\begin{eqnarray}
		\mathcal{F}\left\{\frac{d}{dt}f(t)\right\}(u)
		&=&
		i u\,\mathcal{F}\{f(t)\}(u)
	\end{eqnarray}

	\item \textbf{Multiplication by $t$:}
	\begin{eqnarray}
		\mathcal{F}\{t f(t)\}(u)
		&=&
		i\,\frac{d}{du}\mathcal{F}\{f(t)\}(u)
	\end{eqnarray}
\end{itemize}
			\subsection{Properties of OLCT}\cite{xiang}
\begin{itemize}
	
	\item \textbf{Linearity Property:}
	\begin{eqnarray}
		\mathcal{O}_M\{a f(t) + b g(t)\}(u)
		&=&
		a\,\mathcal{O}_M\{f(t)\}(u) + b\,\mathcal{O}_M\{g(t)\}(u)
	\end{eqnarray}
	
	\item \textbf{Shifting Property:}
	\begin{eqnarray}
		\mathcal{O}_M\{f(t-\tau)\}(u)
		&=&
		e^{(\frac{-iac\tau^2}{2}+ic\tau(u-u_0)+ia\tau\omega_0)}
		\,\mathcal{O}_M\{f(t)\}(u-a\tau)
	\end{eqnarray}
	
	\item \textbf{Differentiation Property:}
	\begin{eqnarray}
		\mathcal{O}_M\left\{\frac{d}{dt}f(t)\right\}(u)=[a\frac{d}{du}-i\frac{ia}{b}du+\frac{ia}{b}(du_0-b\omega_0)-\frac{i}{b}(u_0-u)]\mathcal{O}_M(u)
			\end{eqnarray}
	
	\item \textbf{Multiplication by $t$:}
	\begin{eqnarray}
		\mathcal{O}_M\{t f(t)\}(u)	=[ib\frac{d}{du}+du-(du_0-b\omega_0)]\mathcal{O}_M(u)	
	\end{eqnarray}

    \begin{definition}\cite{hale}
		Delay differential equation:
		  A general
first-order DDE is given by

\begin{eqnarray*}
    \frac{dy}{dx}=f\bigl(t,y(t),y(t-\tau)\bigr),
\end{eqnarray*}

where \(\tau>0\) denotes the delay. To determine a unique solution,
the equation must be supplemented with a history function

\[
y(t)=\phi(t), \qquad t\in[-\tau,0].
\]
	\end{definition}
	
\end{itemize}

	\section{Main Results}
	In this section, we develop an OLCT-based method for solving the considered delay differential equation. The direct OLCT leads to a Volterra-type delay integral equation, which is solved numerically using Brunner's method of steps. An analytical solution is also obtained by exploiting the relationship between the OLCT and the Fourier transform.

	\subsection{Solution of DDE using OLCT}
		Delay differential equation:
	\begin{eqnarray*}
		\frac{d}{dt}X(t)+\alpha tX(t)= \beta X(t-\tau).
	\end{eqnarray*}
	Applying OLCT on both sides we get 
	\begin{eqnarray*}
		\mathcal{O}_M[\frac{d}{dt}X(t)](u)+\alpha\mathcal{O}_M[tX(t)](u)=\beta\mathcal{O}_M[X(t-\tau)](u).
			\end{eqnarray*}
			Using the Properties of OLCT we get 
			\begin{eqnarray*}
				&&[a\frac{d}{du}-\frac{ia}{b}du+\frac{ia}{b}(du_0-b\omega_0)-\frac{i}{b}(u_0-u)]\mathcal{O}_M(u)+\alpha[ib\frac{d}{du}+du-(du_0-b\omega_0)]\mathcal{O}_M(u)\\&=&\beta \mathcal{O}_M[X](u-a\tau)e^{\frac{-iac\tau^2}{2}+ic\tau(u-u_0)+ia\tau\omega_0}
			\end{eqnarray*}
			Rearranging the equation we get 
				\begin{eqnarray*}
					&&(a + \alpha i b)\frac{d}{du}\mathcal{O}_M(u) 
				+ \Bigl[ \alpha d\,(u - u_0) + \alpha b\omega_0 - i c\,(u + u_0) -\frac{2i}{b}u_0 + i a\,\omega_0 \Bigr]\,\mathcal{O}_M(u) \\
				&=& \beta\;\mathcal{O}_M(u - a\tau)\;e^{\left(-\frac{i a c \tau^2}{2} + i c\tau(u - u_0) + i a\tau\omega_0\right)}.
			\end{eqnarray*}
	 Due to the presence of the shifted term $\mathcal{O}_M( u-a \tau)$, the resulting equation is a functional differential equation, which in general does not admit a closed-form analytical solution. Therefore, we proceed by introducing an integrating factor and converting it into an integral equation suitable for iterative solution.

	 we can rewrite the equation,
	 
	  \begin{equation}\label{ordinary diff of olct}
	 	\frac{d}{du}\mathcal{O}_M(u) +\frac{P(u)}{(a+\alpha i b)} \mathcal{O}_M(u)
	 	= \frac{Q(u)}{(a+\alpha i b)}\mathcal{O}_M(u-a\tau),
	 \end{equation}
      Where \begin{eqnarray*}
	 		&& Q(u)=\beta e^{\left(-\frac{iac\tau^2}{2} + ic\tau(u-u_0) + ia\tau\omega_0\right)} and\\ &&P(u)= \Bigl[ \alpha d(u - u_0) + \alpha b\omega_0 - i c(u + u_0) -\frac{2i}{b}u_0 + i a\omega_0 \Bigr].
	\end{eqnarray*}
   Equation \eqref{ordinary diff of olct} is a first-order linear differential equation. Therefore, it can be solved using the integrating factor method.
	 \begin{eqnarray*}
	 	 I.F.=\mu(u) &=&e^{\left[{\int\frac{P(u)}{(a+\alpha i b)} du}\right]}\\
	 	 &=& e^{\left[\frac{1}{(a+\alpha i b)}\int\Bigl[ \alpha d(u - u_0) + \alpha b\omega_0 - i c(u + u_0) -\frac{2i}{b}u_0 + i a\omega_0 \Bigr]du\right]}\\
	 	 &=&e^{\left[\frac{1}{(a+\alpha i b)}\Bigl[ \alpha d(\frac{u^2}{2} - uu_0) + u\alpha b\omega_0 - i c(\frac{u^2}{2} + uu_0) -u\frac{2i}{b}u_0 + ui a\omega_0 \Bigr]\right]}\\
	 \end{eqnarray*} 
	 Multiply I.F. with the equation we get
	 \begin{eqnarray*}
	 	\frac{d}{du}\left[\mu(u)\mathcal{O}_M(u)\right]&=&\mu(u)\frac{Q(u)}{(a+\alpha i b)}\mathcal{O}_M(u-a\tau)
	 \end{eqnarray*}
	Integrating from $0$ to $u$, we get 
	\begin{eqnarray}
		\mu(u)\mathcal{O}_M(u)&=& C+\int_{0}^{u}\mu(u)\frac{Q(u)}{(a+\alpha i b)}\mathcal{O}_M(u-a\tau)du\nonumber\\\mathcal{O}_M(u)&=&\frac{C}{\mu(u)}+\frac{1}{\mu(u)}\int_{0}^{u}\frac{Q(s)}{(a+\alpha i b)}\mu(s)\mathcal{O}_M(s-a\tau)ds \label{volterra with delay}
	\end{eqnarray}
	This is a Volterra-type integral equation with delay\cite{burton}. 
Now, if $a\tau \neq 0$, to solve equation \eqref{volterra with delay}, we use Brunner’s analysis\cite{brunner, blom} for delay Volterra equations, which admits a unique solution that can be constructed iteratively using the method of steps.

	 	\subsection{Numerical Solution via Method of Steps}
	 	
	 	Let $h = a\tau > 0$. Due to the presence of delay, we define the initial function.

        For the numerical computations presented in this work, we choose

\begin{eqnarray*}
    \phi(u)=\frac{C}{\mu(u)}, \qquad u\in[-h,0].
\end{eqnarray*}

Consequently,

\begin{eqnarray*}
    \mathcal{O}_M(u)=\frac{C}{\mu(u)},
\qquad u\in[-h,0].
\end{eqnarray*}
	\subsubsection*{Let $0 \le u \le h$}
	 	
	 	For $u \in [0,h]$, we have $u-h \in [-h,0]$, hence
	 	\begin{eqnarray*}
	 		\mathcal{O}_M(s-h)=\phi(s-h).
	 	\end{eqnarray*}
	 	
	 	Thus, the integral equation reduces to
	 	\begin{eqnarray*}
	 		\mathcal{O}_M(u)=\frac{C}{\mu(u)}+\frac{1}{\mu(u)}\int_{0}^{u}K(s)\phi(s-h)\,ds,
	 	\end{eqnarray*} where 
	 	\begin{eqnarray*}
	 	K(s)= \frac{Q(s)}{(a+\alpha i b)}\mu(s)
	 	\end{eqnarray*}
	 	which is a standard Volterra integral equation without delay.
	 	
	 	\subsubsection*{Now $h \le u \le 2h$}
	 	
	 	For this interval, the delay term is given by
	 	\begin{eqnarray*}
	 		\mathcal{O}_M(s-h)=
	 		\begin{cases}
	 			\phi(s-h), & 0 \le s < h, \\
	 			\mathcal{O}_M(s-h), & h \le s \le u.
	 		\end{cases}
	 	\end{eqnarray*}
	 	
	 	Hence,
	 	\begin{eqnarray*}
	 		\mathcal{O}_M(u)=\frac{C}{\mu(u)}+\frac{1}{\mu(u)}\left[
	 		\int_{0}^{h}K(s)\phi(s-h)\,ds
	 		+\int_{h}^{u}K(s)\mathcal{O}_M(s-h)\,ds
	 		\right].
	 	\end{eqnarray*}
	 	
	 	Proceeding iteratively, for $u \in [nh,(n+1)h]$, the solution is constructed using previously computed values:
	 	\begin{eqnarray}
	 		\mathcal{O}_M(u)=\frac{C}{\mu(u)}+\frac{1}{\mu(u)}\int_{0}^{u}K(s)\mathcal{O}_M(s-h)\,ds.\label{olct domain solution}
	 	\end{eqnarray}
	 	
	 	This procedure is known as the method of steps and allows the construction of the solution over successive intervals.
 \begin{figure}[h]
\centering
\begin{tabular}{cc}
\includegraphics[width=5.5cm]{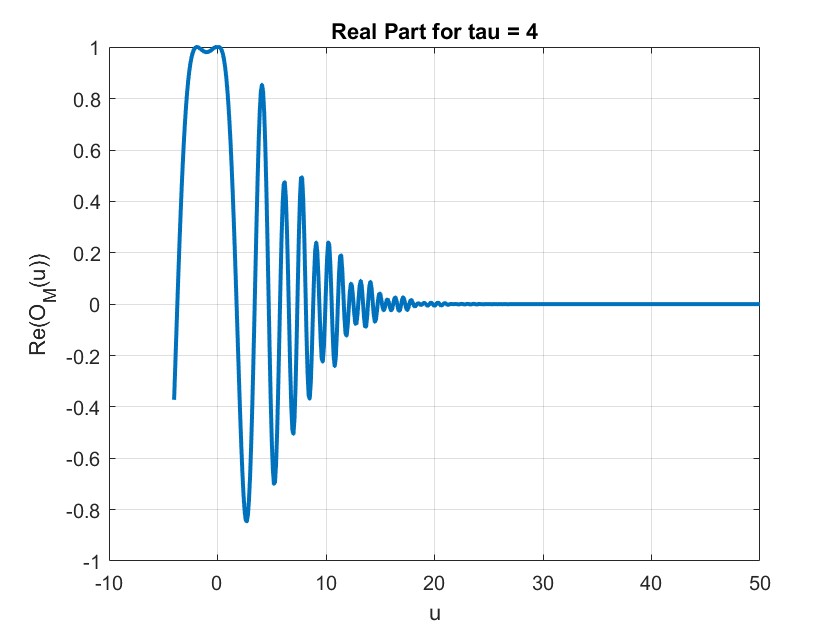} &
\includegraphics[width=5.5cm]{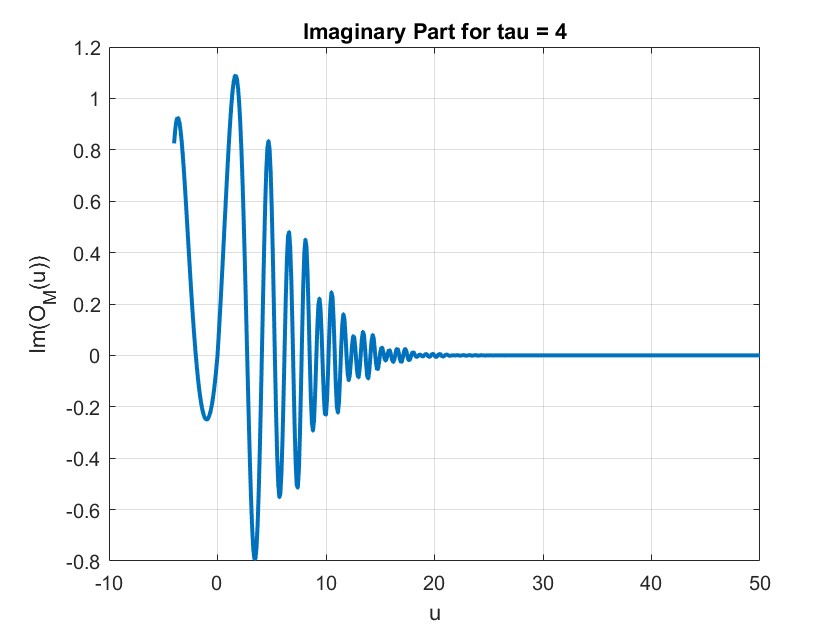} \\

(a) Real part at $\tau=4$, $\beta=-4$, $\alpha=10$ &
(b) Imaginary part at $\tau=4$, $\beta=-4$, $\alpha=10$ \\[0.3cm]

\includegraphics[width=5.5cm]{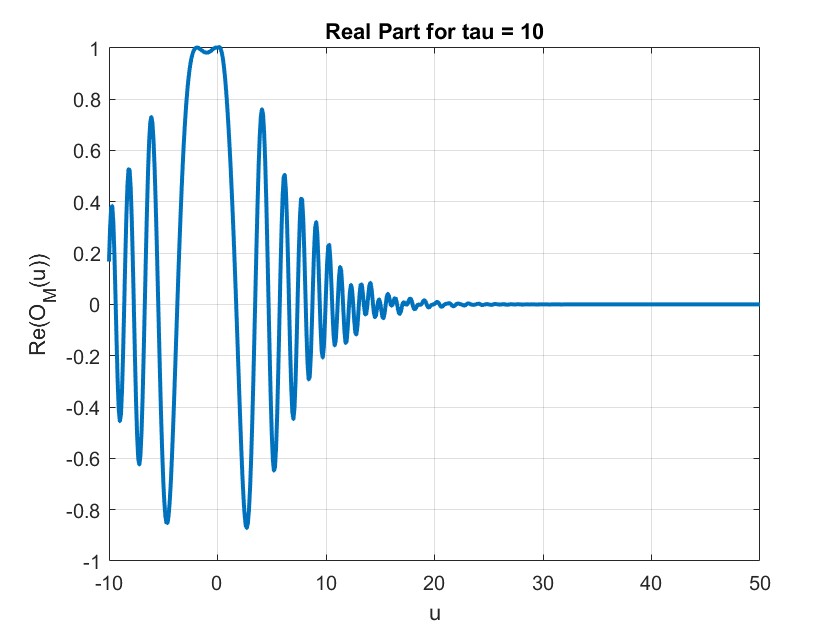} &
\includegraphics[width=5.5cm]{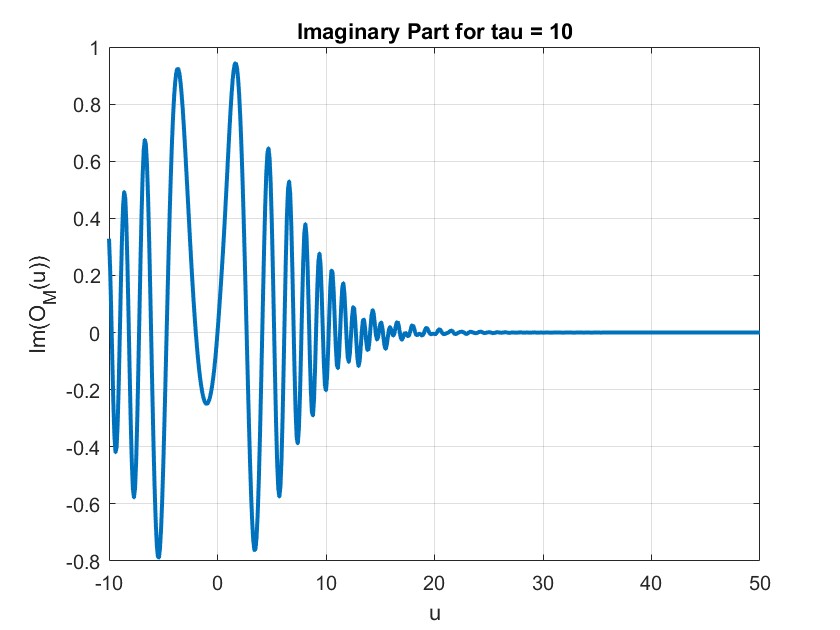} \\

(c) Real part at $\tau=10$, $\beta=-4$, $\alpha=10$ &
(d) Imaginary part at $\tau=10$, $\beta=-4$, $\alpha=10$ \\[0.3cm]

\includegraphics[width=5.5cm]{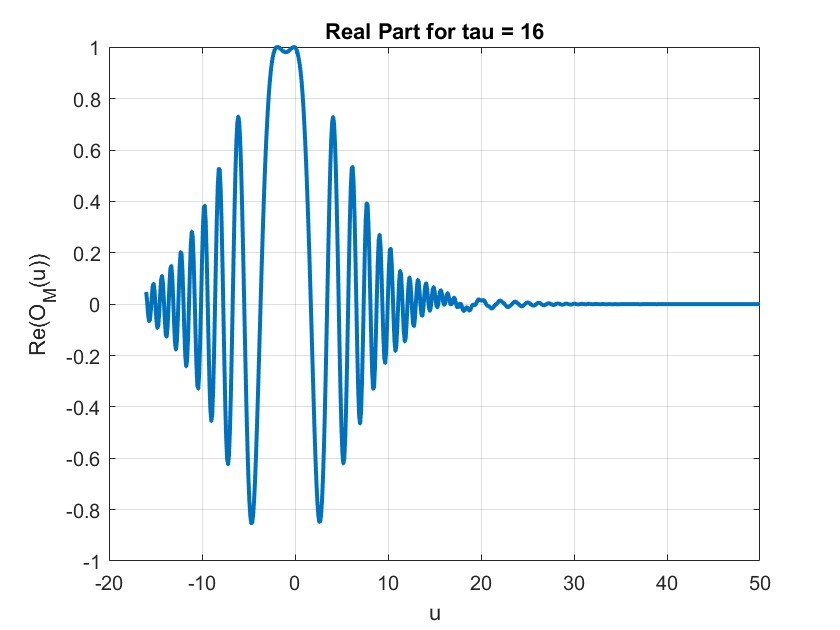} &
\includegraphics[width=5.5cm]{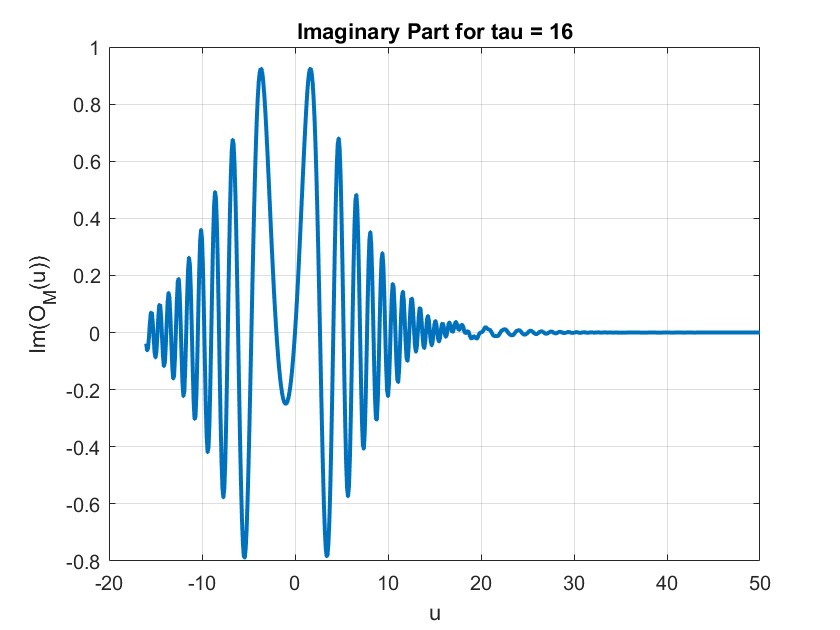} \\

(e) Real part at $\tau=16$, $\beta=-4$, $\alpha=10$ &
(f) Imaginary part at $\tau=16$, $\beta=-4$, $\alpha=10$
\end{tabular}

\caption{Real and imaginary parts of the solution of equation \eqref{olct domain solution} for $\beta=-4$ and $\alpha=10$ corresponding to different values of the delay parameter $\tau$.}
\label{fig:dde_sol_olct_negative}
\end{figure}
   
\begin{table}[htbp]
\centering
\caption{Values of $\mathrm{Re}(\mathcal{O}_M(u))$ and $\mathrm{Im}(\mathcal{O}_M(u))$ for different values of $\tau$.(negative $\beta$)}
\label{tab:OM_values}
\begin{tabular}{|c|c|c|c|c|c|c|}
\hline
& \multicolumn{2}{c|}{$\tau = 4$}
& \multicolumn{2}{c|}{$\tau = 10$}
& \multicolumn{2}{c|}{$\tau = 16$} \\

$u$
& $\mathrm{Re}(\mathcal{O}_M(u))$
& $\mathrm{Im}(\mathcal{O}_M(u))$
& $\mathrm{Re}(\mathcal{O}_M(u))$
& $\mathrm{Im}(\mathcal{O}_M(u))$
& $\mathrm{Re}(\mathcal{O}_M(u))$
& $\mathrm{Im}(\mathcal{O}_M(u))$ \\
\hline
4  &  0.829533 & -0.179392 &  0.734535 & -0.220401 &  0.709742 & -0.222034 \\
8  &  0.259186 &  0.396731 &  0.204982 &  0.348096 &  0.179075 &  0.324459 \\
12 & -0.082486 & -0.072855 & -0.091615 & -0.086283 & -0.074590 & -0.090715 \\
16 & -0.024531 &  0.001505 & -0.003923 &  0.011226 & -0.002262 &  0.018301 \\
20 & -0.006544 & -0.000284 &  0.001518 & -0.005396 &  0.013523 &  0.000702 \\
\hline
\end{tabular}
 
\end{table}
 
	\begin{figure}[h]
\centering
\begin{tabular}{cc}
\includegraphics[width=5.5cm]{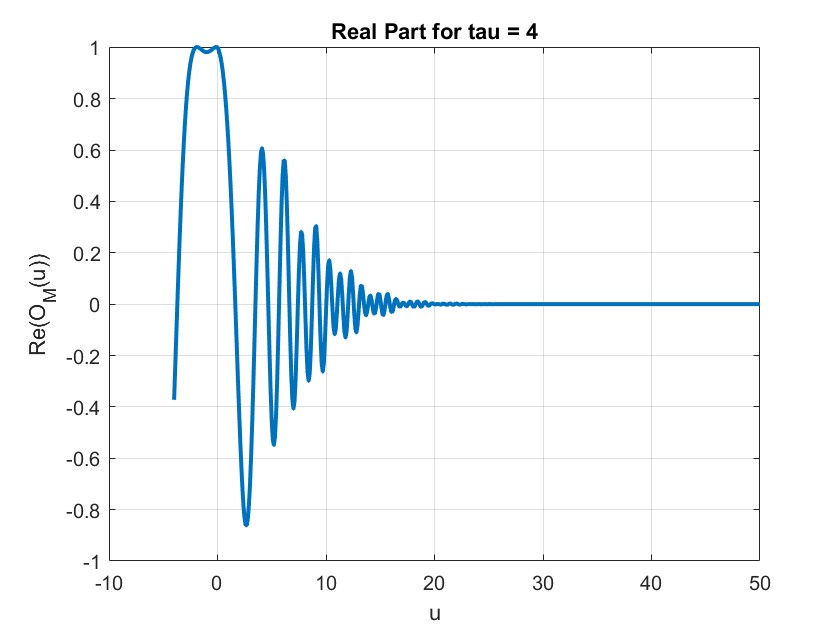} &
\includegraphics[width=5.5cm]{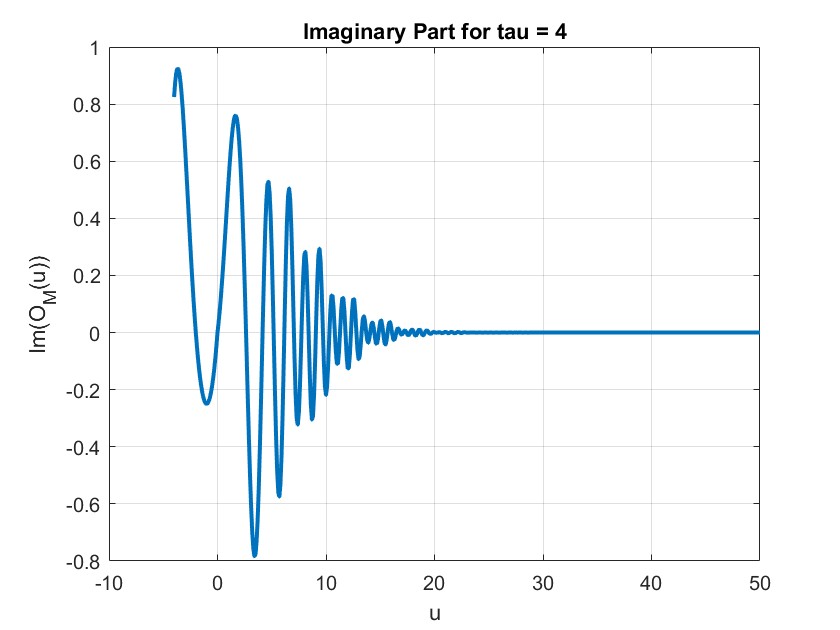} \\

(a) Real part at $\tau=4$, $\beta=4$, $\alpha=10$ &
(b) Imaginary part at $\tau=4$, $\beta=4$, $\alpha=10$ \\[0.3cm]

\includegraphics[width=5.5cm]{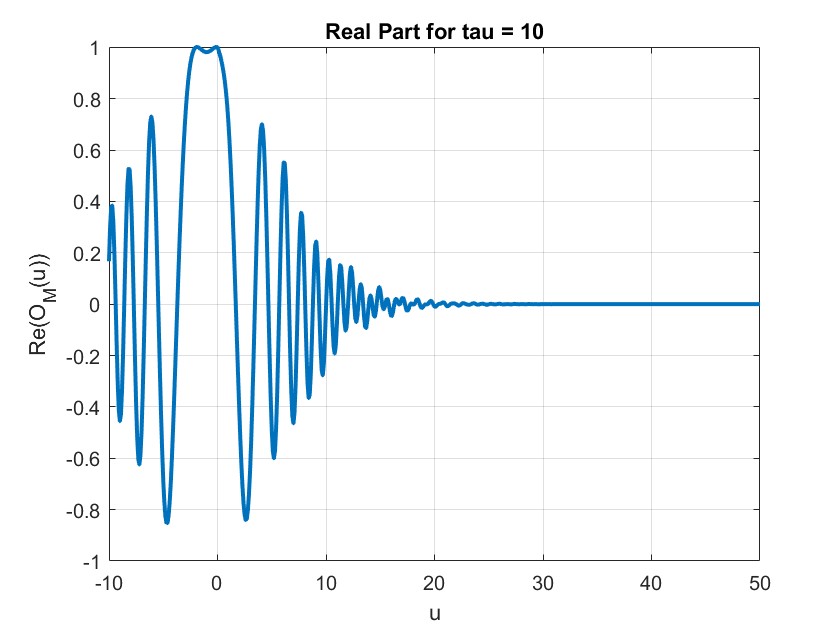} &
\includegraphics[width=5.5cm]{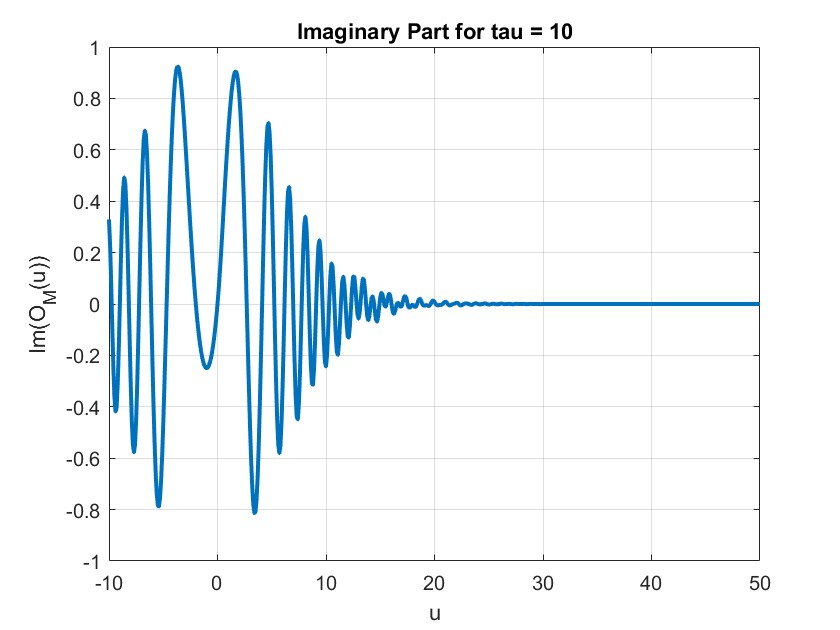} \\

(c) Real part at $\tau=10$, $\beta=4$, $\alpha=10$ &
(d) Imaginary part at $\tau=10$, $\beta=4$, $\alpha=10$ \\[0.3cm]

\includegraphics[width=5.5cm]{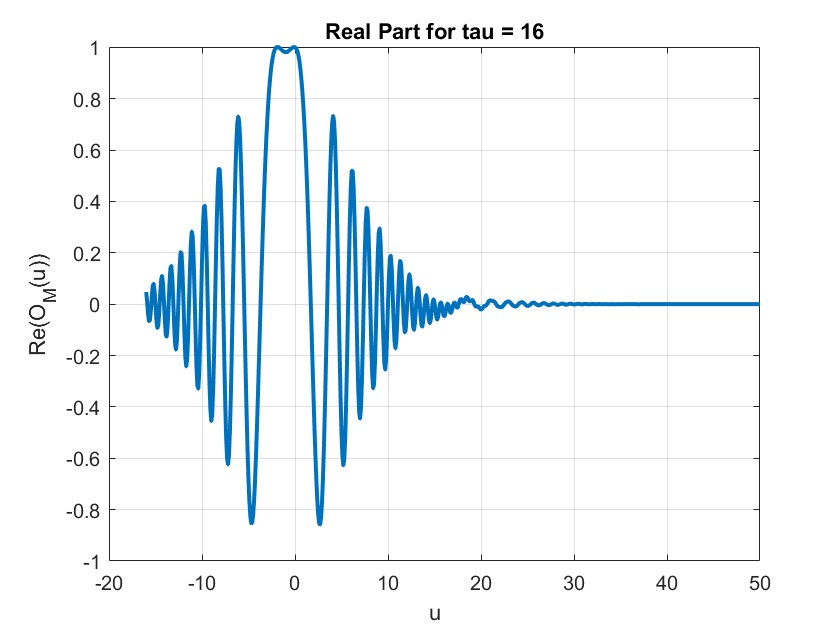} &
\includegraphics[width=5.5cm]{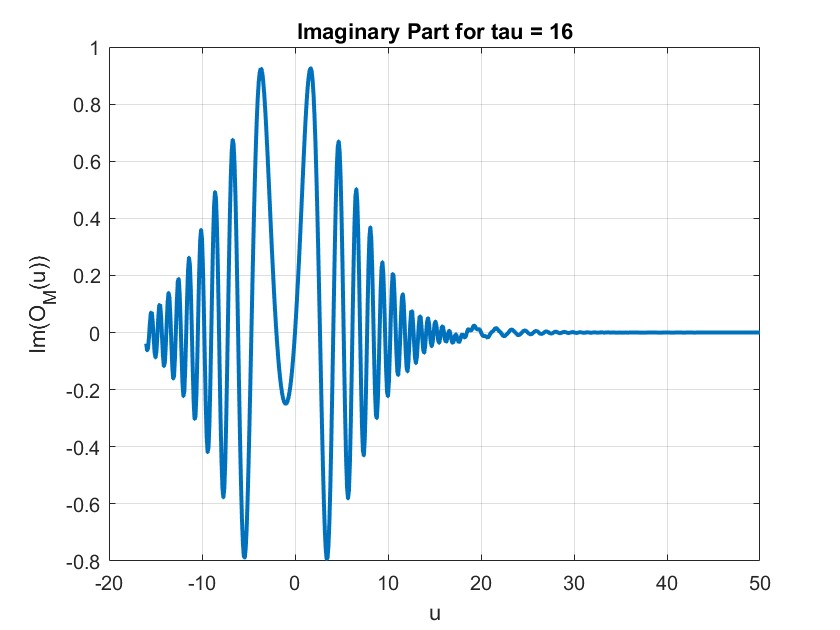} \\

(e) Real part at $\tau=16$, $\beta=4$, $\alpha=10$ &
(f) Imaginary part at $\tau=16$, $\beta=4$, $\alpha=10$
\end{tabular}

\caption{Real and imaginary parts of the solution of equation \eqref{olct domain solution} for $\beta=4$ and $\alpha=10$ corresponding to different values of the delay parameter $\tau$.}
\label{fig:dde_sol_olct_positive}
\end{figure}

    \begin{table}[htbp]
\centering
\caption{Values of $\operatorname{Re}(\mathcal{O}_M)$ and $\operatorname{Im}(\mathcal{O}_M)$ for different $\tau$ values .}
\begin{tabular}{|r|rr|rr|rr|}
\hline
& \multicolumn{2}{c|}{$\tau=4$}
& \multicolumn{2}{c|}{$\tau=10$}
& \multicolumn{2}{c|}{$\tau=16$} \\
$u$
& $\operatorname{Re}(OM)$ & $\operatorname{Im}(OM)$
& $\operatorname{Re}(OM)$ & $\operatorname{Im}(OM)$
& $\operatorname{Re}(OM)$ & $\operatorname{Im}(OM)$ \\
\hline

  4 &  0.587797 & -0.256165 &  0.682795 & -0.215156 &  0.707588 & -0.213523 \\
  8 &  0.088192 &  0.269954 &  0.129215 &  0.309445 &  0.155122 &  0.333082 \\
 12 & -0.039102 & -0.126040 & -0.028800 & -0.126181 & -0.046937 & -0.123400 \\
 16 & -0.028450 &  0.020903 & -0.047230 &  0.011472 & -0.047715 &  0.004995 \\
 20 & -0.000871 &  0.001690 & -0.009159 &  0.008419 & -0.021571 &  0.001125 \\
\hline
\end{tabular}
 
\label{tab:OM_tau_comparison}
\end{table}

\begin{table}[htbp]
\centering
\caption{Effect of parameters on the OLCT-domain solution $\mathcal{O}_M(u)$.}
\label{tab:parameter_effects}
\begin{tabular}{|c|p{9cm}|}
\hline
\textbf{Parameter} & \textbf{Observed effect on the solution} \\
\hline
$\tau$ (delay) &
Controls how much the past values influence the present solution. As $\tau$ increases, the effect of the delayed term becomes more noticeable and the shape of the solution changes. \\
\hline
$\beta>0$ &
The delayed term supports the present solution. The oscillations remain visible for a longer interval and the decay is slower. \\
\hline
$\beta<0$ &
The delayed term opposes the present solution. The oscillations decrease more quickly and the solution approaches zero faster. \\
\hline
$\alpha$ &
Controls the damping of the solution. Larger values of $\alpha$ reduce the amplitude more rapidly and lead to faster decay. \\
\hline
\end{tabular}
\end{table}

	  Next, applying the inverse OLCT to $\mathcal{O}_M(u)$, we obtain 
	\begin{eqnarray}
		f(t) = \frac{1}{\sqrt{2\pi(-i)b}}
		\int_{\mathbb{R}}
		\mathcal{K}_{M^{-1}}(t,u)\,
		\mathcal{O}_M(u)\,du.
	\end{eqnarray}
	 Substituting the expression of $\mathcal{O}_M(u)$ from \eqref{volterra with delay}, we get
	 \begin{eqnarray*}
	 	f(t)
	 	&=&
	 	\frac{1}{\sqrt{2\pi(-i)b}}
	 	\int_{\mathbb{R}}
	 \mathcal{K}_{M^{-1}}(t,u)\,
	 	\left[
	 	\frac{C}{\mu(u)}
	 	+
	 	\frac{1}{\mu(u)}
	 	\int_{0}^{u}
	 	K(s)\mathcal{O}_M(s-h)\,ds
	 	\right] du.
	 \end{eqnarray*}
	 
	Substituting the expressions of $\mu(u)$ and $K(s)$ into the inverse OLCT representation, we obtain
	\begin{eqnarray*}
		f(t)&=&	\frac{1}{\sqrt{2\pi(-i)b}}
		\int_{\mathbb{R}}	e^{\Phi(t,u)}\Bigg[C+\frac{\beta}{(a+\alpha ib)}
\int_{0}^{u}e^{\Theta(u,s)}\mathcal{O}_M(s-h)\,ds\Bigg]du,
	\end{eqnarray*}
	where
	\begin{eqnarray*}
		\Phi(t,u)
		&=&
		\frac{-i}{2b}
		\Big(
		a t^2+2t(u_0-u)
		-2u(du_0-b\omega_0)
		+du^2+du_0^2
		\Big) \\
		&&-
		\frac{1}{(a+\alpha ib)}
		\Big[
		\alpha d\Big(\frac{u^2}{2}-uu_0\Big)
		+u\alpha b\omega_0
		-ic\Big(\frac{u^2}{2}+uu_0\Big)
		-\frac{2i}{b}u_0u
		+ia\omega_0u
		\Big],
	\end{eqnarray*}
	and
	\begin{eqnarray*}
		\Theta(u,s)
		&=&
		-\frac{iac\tau^2}{2}
		+ic\tau(s-u_0)
		+ia\tau\omega_0 \\
		&&+
		\frac{1}{(a+\alpha ib)}
		\Big[
		\alpha d\Big(\frac{s^2}{2}-su_0\Big)
		+s\alpha b\omega_0
		-ic\Big(\frac{s^2}{2}+su_0\Big)
		-\frac{2i}{b}u_0s
		+ia\omega_0s
		\Big].
	\end{eqnarray*}
	
	Thus, the inverse OLCT representation consists of an explicit kernel term together with a delay-dependent integral term involving the history function $\mathcal{O}_M(s-h)$.
	 
	 Since $\mathcal{O}_M(s-h)$ depends on previous values of the solution, it is difficult to obtain an exact closed-form solution. Therefore, the solution is computed numerically using the method of steps together with numerical integration for the inverse OLCT.
	 
	 	\subsection{Special Case $a=0$ }
	 	
	 	Starting from the Volterra integral equation with delay,
	 	\begin{eqnarray}
	 		\mathcal{O}_M(u)
	 		&=&
	 		\frac{C}{\mu(u)}
	 		+
	 		\frac{1}{\mu(u)}
	 		\int_{0}^{u}
	 		\frac{Q(s)}{(a+\alpha i b)}\mu(s)\mathcal{O}_M(s-a\tau)\,ds 		
	 	\end{eqnarray}
	 	
	 	\noindent
	 	For the special case $a=0$, we have
	 	\[
	 	a+\alpha i b = \alpha i b, \qquad s-a\tau = s.
	 	\]
	 	
	 	Also, using the OLCT condition $ad-bc=1$, we obtain $c=-\frac{1}{b}$. Hence,
	 	\begin{eqnarray*}
	 		Q(s)
	 		&=&
	 		\beta e^{\left(ic\tau(s-u_0)\right)}
	 		=
	 		\beta e^{\left(-i\frac{\tau}{b}(s-u_0)\right)}.
	 	\end{eqnarray*}
	 	
	 	Substituting into equation \eqref{volterra with delay}, we get
	 	\begin{eqnarray}
	 		\mathcal{O}_M(u)
	 		&=&
	 		\frac{C}{\mu(u)}
	 		+
	 		\frac{\beta}{\alpha i b\,\mu(u)}
	 		\int_{0}^{u}
	 		\mu(s)\,
	 		e^{-i\frac{\tau}{b}(s-u_0)}
	 		\mathcal{O}_M(s)\,ds.
	 	\end{eqnarray}
	 	
	 	\noindent
	 	Multiplying both sides by $\mu(u)$, we obtain
	 	\begin{eqnarray}
	 		\mu(u)\mathcal{O}_M(u)
	 		&=&
	 		C
	 		+
	 		\frac{\beta}{\alpha i b}
	 		\int_{0}^{u}
	 		e^{-i\frac{\tau}{b}(s-u_0)}
	 		\mu(s)\mathcal{O}_M(s)\,ds.
	 	\end{eqnarray}
	 	
	 	Let
	 	\[
	 	Y(u)=\mu(u)\mathcal{O}_M(u),
	 	\]
	 	then
	 	\begin{eqnarray}
	 		Y(u)
	 		&=&
	 		C
	 		+
	 		\frac{\beta}{\alpha i b}
	 		\int_{0}^{u}
	 		e^{-i\frac{\tau}{b}(s-u_0)}Y(s)\,ds.
	 	\end{eqnarray}
	 	
	 	Differentiating both sides with respect to $u$, we get
	 	\begin{eqnarray}
	 		\frac{dY}{du}
	 		&=&
	 		\frac{\beta}{\alpha i b}
	 		e^{-i\frac{\tau}{b}(u-u_0)}Y(u).
	 	\end{eqnarray}
	 	
	 	Separating variables,
	 	\begin{eqnarray}
	 		\frac{dY}{Y}
	 		&=&
	 		\frac{\beta}{\alpha i b}
	 		e^{-i\frac{\tau}{b}(u-u_0)}du.
	 	\end{eqnarray}
	 	
	 	Integrating,
	 	\begin{eqnarray}
	 		\ln Y
	 		&=&
	 		\frac{\beta}{\alpha i b}
	 		\int e^{-i\frac{\tau}{b}(u-u_0)}du.
	 	\end{eqnarray}
	 	
	 	Evaluating the integral,
	 	\begin{eqnarray*}
	 		\int e^{-i\frac{\tau}{b}(u-u_0)}du
	 		&=&
	 		-\frac{b}{i\tau}
	 		e^{-i\frac{\tau}{b}(u-u_0)}.
	 	\end{eqnarray*}
	 	
	 	Thus,
	 	\begin{eqnarray}
	 		\ln Y
	 		&=&
	 		\frac{\beta}{\alpha\tau}
	 		e^{-i\frac{\tau}{b}(u-u_0)}.
	 	\end{eqnarray}
	 	
	 	Hence,
	 	\begin{eqnarray}
	 		Y(u)
	 		&=&
	 		C
	 		\exp\left(
	 		\frac{\beta}{\alpha\tau}
	 		e^{-i\frac{\tau}{b}(u-u_0)}
	 		\right).
	 	\end{eqnarray}
	 	
	 	Therefore,
	 	\begin{eqnarray}
	 		\mathcal{O}_M(u)
	 		&=&
	 		\frac{C}{\mu(u)}
	 		\exp\left(
	 		\frac{\beta}{\alpha\tau}
	 		e^{-i\frac{\tau}{b}(u-u_0)}
	 		\right).
	 	\end{eqnarray}
	 	Using the inverse OLCT formula,
	 	\begin{eqnarray}
	 		X(t)
	 		&=&
	 		\frac{1}{\sqrt{2\pi(-i)b}}
	 		\int_{\mathbb{R}}
	 		e^{\frac{-i}{2b}\left(a t^2 + 2t(u_0-u)-2u(du_0-b\omega_0)+du^2+du_0^2\right)}
	 		\mathcal{O}_M(u)\,du.
	 	\end{eqnarray}
	 	
	 	For $a=0$, this reduces to
	 	\begin{eqnarray}
	 		X(t)
	 		&=&
	 		\frac{1}{\sqrt{2\pi(-i)b}}
	 		\int_{\mathbb{R}}
	 		e^{\frac{-i}{2b}\left(2t(u_0-u)-2u(du_0-b\omega_0)+du^2+du_0^2\right)}
	 		\mathcal{O}_M(u)\,du.
	 	\end{eqnarray}
	 	
	 	Substituting $\mathcal{O}_M(u)$, we get
	 	\begin{eqnarray}
	 		X(t)
	 		&=&
	 		\frac{C}{\sqrt{2\pi(-i)b}}
	 		\int_{\mathbb{R}}
	 		e^{\frac{-i}{2b}\left(2t(u_0-u)-2u(du_0-b\omega_0)+du^2+du_0^2\right)}
	 		\frac{1}{\mu(u)} \nonumber\\
	 		&& \times
	 		\exp\left(
	 		\frac{\beta}{\alpha\tau}
	 		e^{-i\frac{\tau}{b}(u-u_0)}
	 		\right)
	 		du. 
	 	\end{eqnarray}
	 	
	 	After simplification, we obtain
	 	\begin{eqnarray}
	 		X(t)
	 		&=&
	 		\frac{C}{\sqrt{2\pi b}}
	 		e^{i\left(\frac{\pi}{4}-\frac{1}{2b}(2tu_0+du_0^2)\right)}
	 		\int_{\mathbb{R}}
	 		\exp\left(
	 		-\frac{b}{2\alpha}u^2
	 		+\frac{i t}{b}u
	 		+\frac{\beta}{\alpha\tau}
	 		e^{-i\frac{\tau}{b}(u-u_0)}
	 		\right)du.\label{special case}
	 	\end{eqnarray}
\begin{figure}[h]
\centering
\begin{tabular}{cc}

\includegraphics[width=5.5cm]{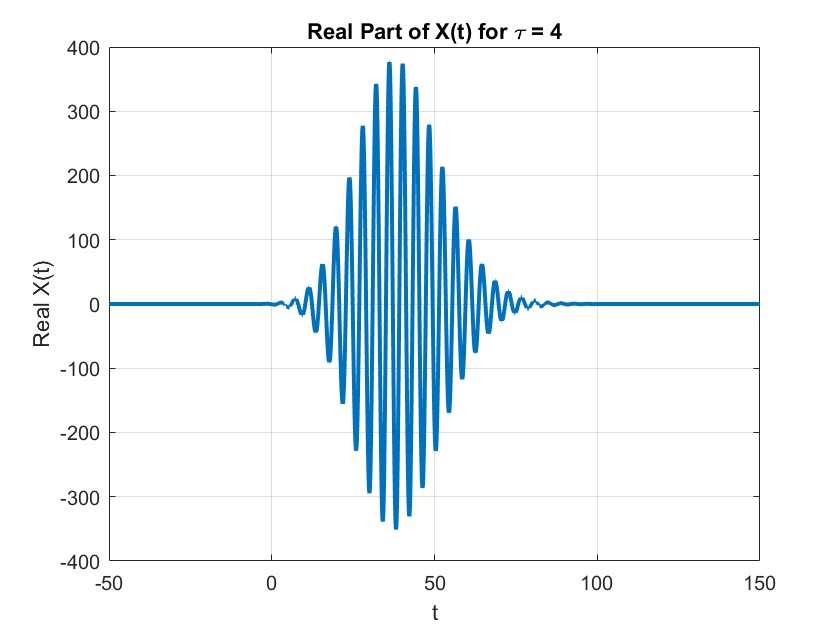} &
\includegraphics[width=5.5cm]{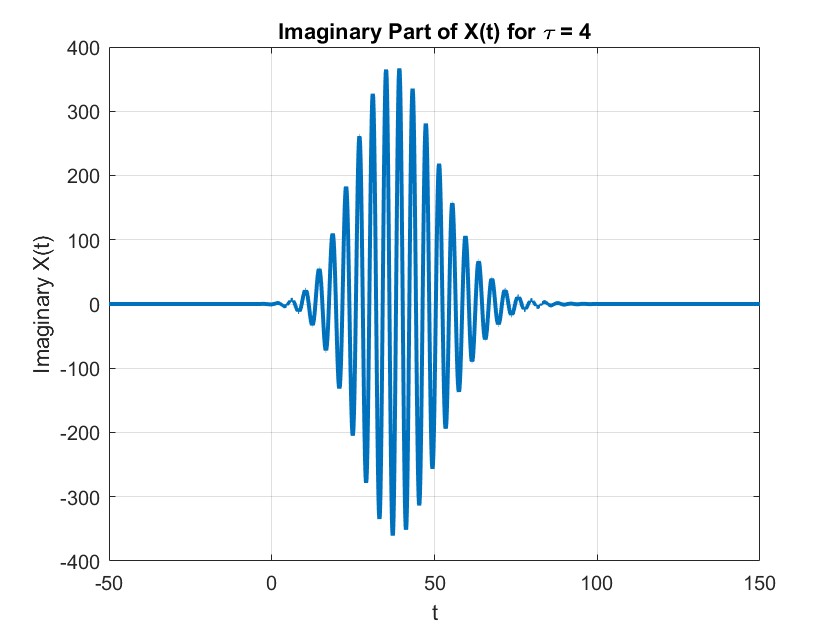} \\

(a) Real part at $\tau=4$, $\beta=-4$, $\alpha=0.1$ &
(b) Imaginary part at $\tau=4$, $\beta=-4$, $\alpha=0.1$ \\[0.3cm]

\includegraphics[width=5.5cm]{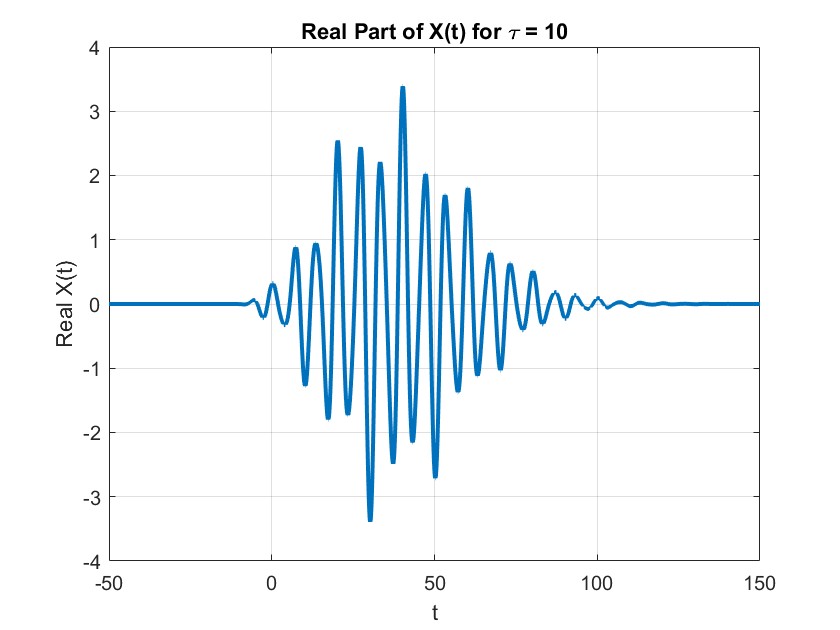} &
\includegraphics[width=5.5cm]{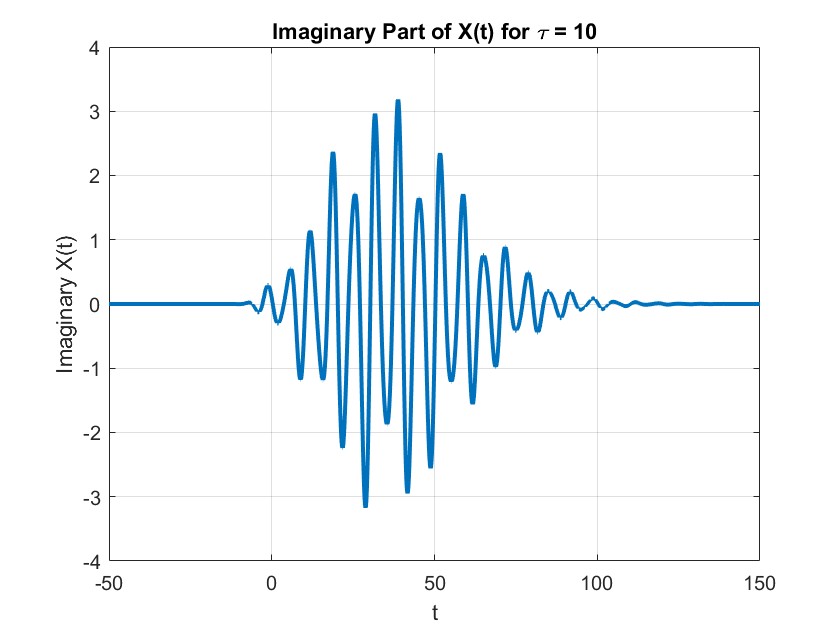} \\

(c) Real part at $\tau=10$, $\beta=-4$, $\alpha=1$ &
(d) Imaginary part at $\tau=10$, $\beta=-4$, $\alpha=0.1$ \\[0.3cm]

\includegraphics[width=5.5cm]{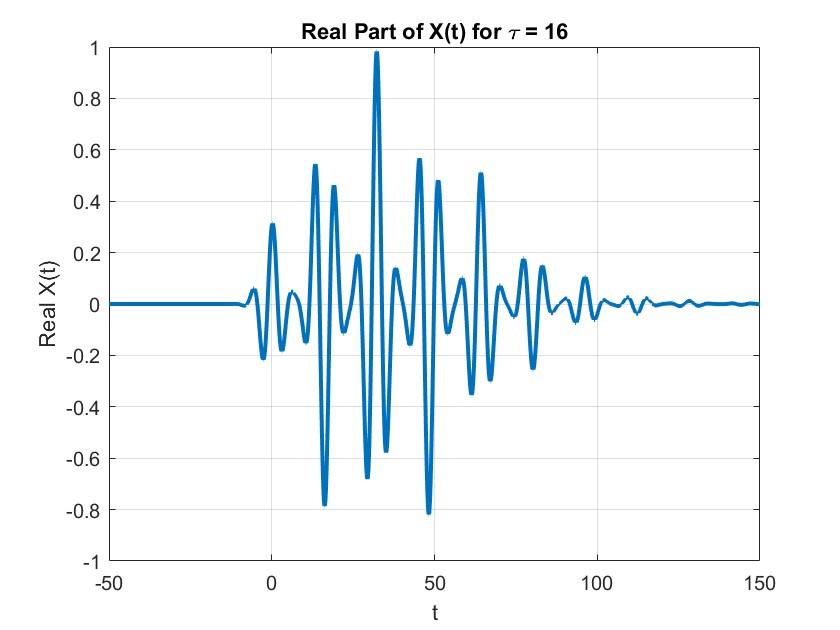} &
\includegraphics[width=5.5cm]{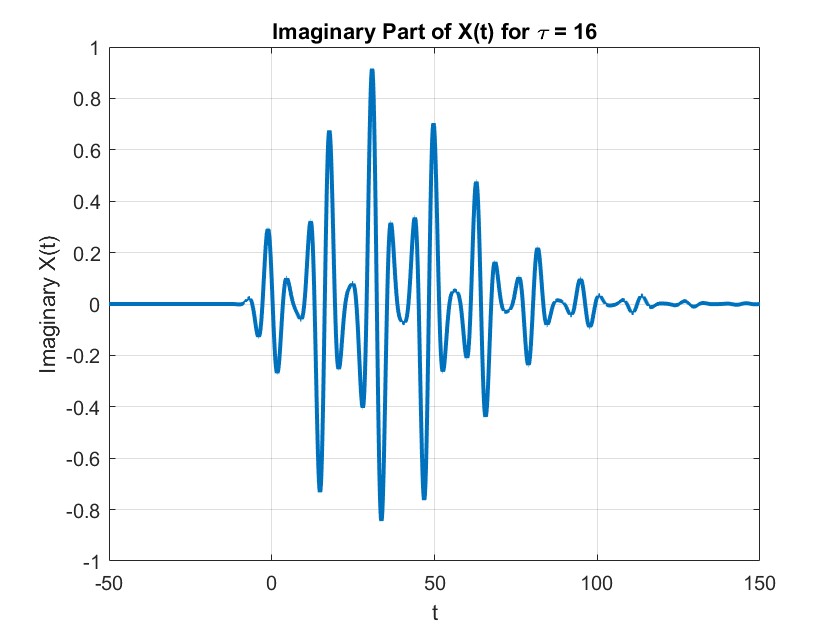} \\

(e) Real part at $\tau=16$, $\beta=-4$, $\alpha=0.1$ &
(f) Imaginary part at $\tau=16$, $\beta=-4$, $\alpha=0.1$

\end{tabular}

\caption{Real and imaginary parts of the solution obtained for the Fourier special case ($a=0$) with $\beta=-4$ corresponding to different values of the delay parameter $\tau$ for solution \eqref{special case}.}
\label{fig:fourier_special_case_negative}

\end{figure}
         
        \begin{table}[h]
\centering
\caption{Numerical values for different values of $\tau$ for solution \eqref{special case} (negative $\beta$)}
\begin{tabular}{|c|c|c|c|}

\hline
$t$ & Real Part & Imag Part & Difference \\ 
\hline

\multicolumn{4}{|c|}{$\tau = 4$} \\ 
\hline
0.00  & 0.584264   & 1.978525   & 1.394261 \\
4.00  & -5.102500  & -9.258305  & 4.155805 \\
8.00  & 23.739020  & 30.156555  & 6.417535 \\
12.00 & -76.270423 & -74.595036 & 1.675387 \\
16.00 & 187.517692 & 147.785408 & 39.732285 \\
20.00 & -373.371029 & -242.979722 & 130.391306 \\
\hline

\multicolumn{4}{|c|}{$\tau = 10$} \\ 
\hline
0.00  & 0.308993  & 0.095493  & 0.213500 \\
4.00  & -0.328488 & 0.076570  & 0.405058 \\
8.00  & 0.682266  & -0.795550 & 1.477816 \\
12.00 & 0.105450  & 1.118782  & 1.013332 \\
16.00 & -0.646920 & -1.147429 & 0.500510 \\
20.00 & 2.450475  & 0.727042  & 1.723433 \\
\hline

\multicolumn{4}{|c|}{$\tau = 16$} \\ 
\hline
0.00  & 0.303438  & 0.089032  & 0.214406 \\
4.00  & -0.119962 & 0.077200  & 0.197162 \\
8.00  & 0.015266  & -0.042038 & 0.057304 \\
12.00 & 0.147266  & 0.323510  & 0.176243 \\
16.00 & -0.758594 & -0.222578 & 0.536016 \\
20.00 & 0.299196  & -0.192795 & 0.491990 \\
\hline
\end{tabular}

\end{table}

\begin{figure}[h]
\centering
\begin{tabular}{cc}
\includegraphics[width=5.5cm]{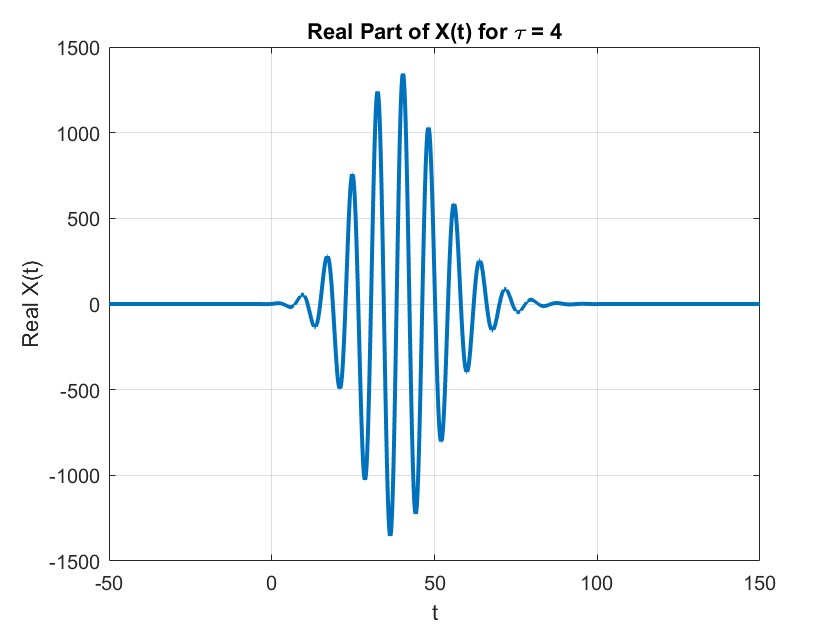} &
\includegraphics[width=5.5cm]{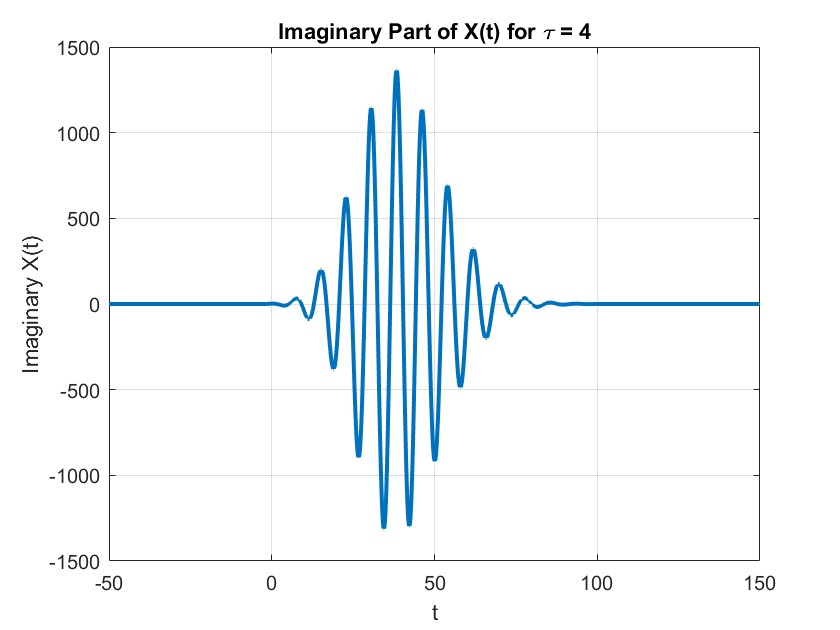} \\

(a) Real part at $\tau=4$, $\beta=4$, $\alpha=0.1$ &
(b) Imaginary part at $\tau=4$, $\beta=4$, $\alpha=0.1$ \\[0.3cm]

\includegraphics[width=5.5cm]{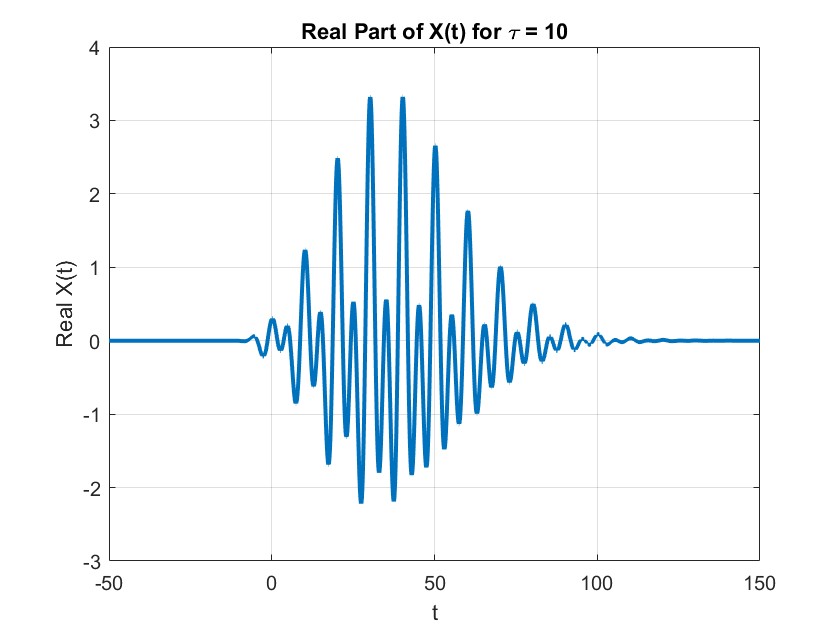} &
\includegraphics[width=5.5cm]{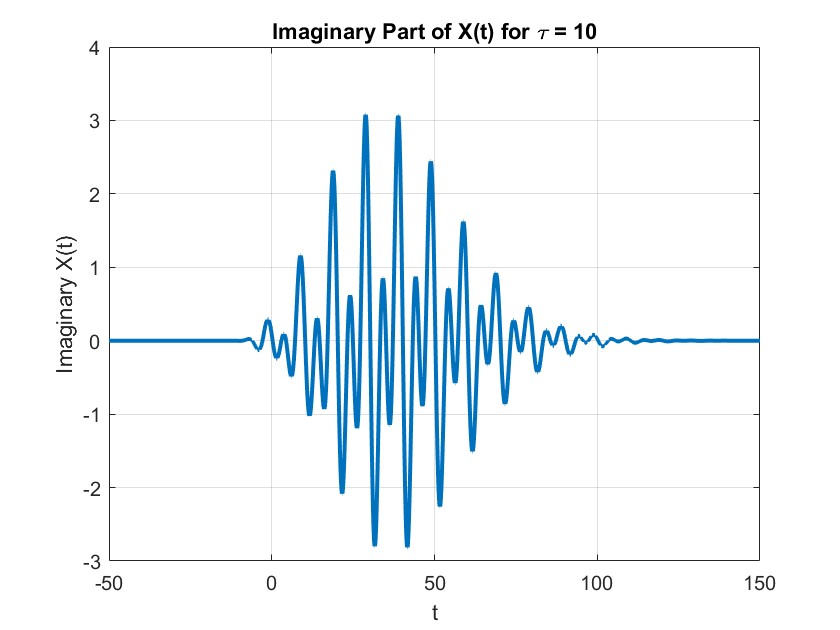} \\

(c) Real part at $\tau=10$, $\beta=4$, $\alpha=0.1$ &
(d) Imaginary part at $\tau=10$, $\beta=4$, $\alpha=0.1$ \\[0.3cm]

\includegraphics[width=5.5cm]{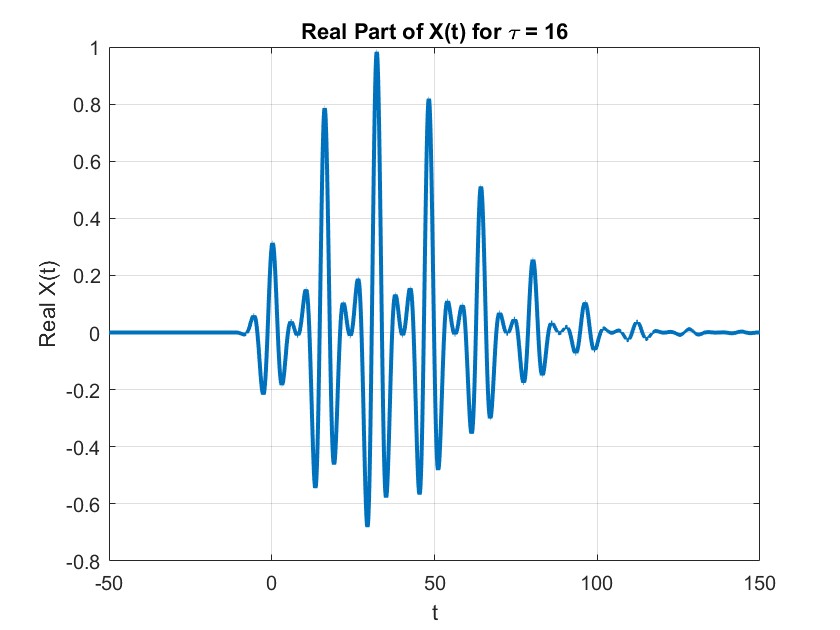} &
\includegraphics[width=5.5cm]{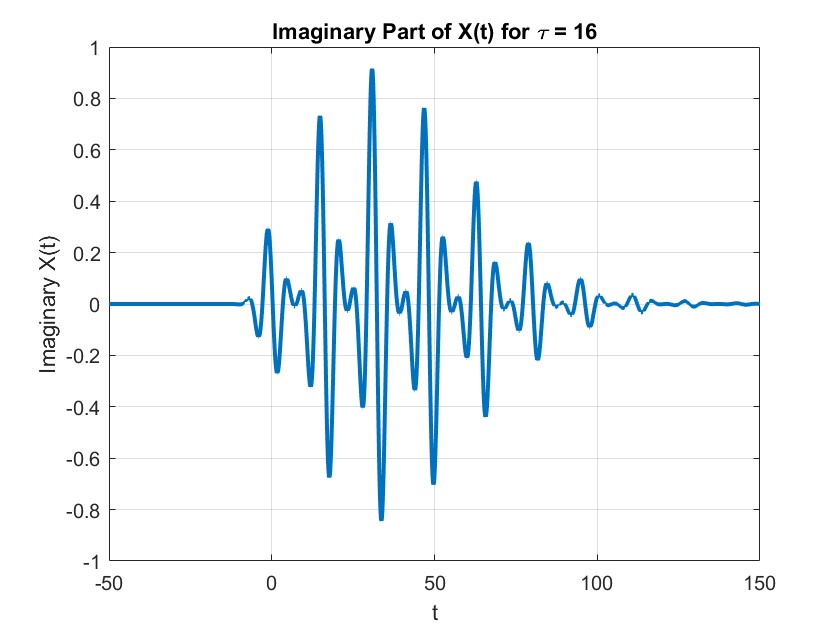} \\

(e) Real part at $\tau=16$, $\beta=4$, $\alpha=0.1$ &
(f) Imaginary part at $\tau=16$, $\beta=4$, $\alpha=0.1$
\end{tabular}

\caption{Real and imaginary parts of the solution obtained for the Fourier special case ($a=0$) with $\beta=4$ and $\alpha=0.1$ corresponding to different values of the delay parameter $\tau$ for solution \eqref{special case}.}
\label{fig:fourier_special_case_positive}
\end{figure}
        \begin{table}[h]
\centering
\caption{Numerical values for different values of $\tau$ for solution \eqref{special case} (positive $\beta$)}
\begin{tabular}{|c|c|c|c|}
\hline
$t$ & Real Part & Imag Part & Difference \\ 
\hline
\multicolumn{4}{|c|}{$\tau = 4$} \\ 
\hline
0.00  & -0.517012 & -0.629953 & 0.112941 \\
4.00  & -0.831699 & -3.575463 & 2.743763 \\
8.00  & 2.114390  & -11.522787 & 13.637177 \\
12.00 & 15.924378 & -25.660236 & 41.584614 \\
16.00 & 50.548747 & -42.346902 & 92.895649 \\
20.00 & 111.559575 & -52.388458 & 163.948033 \\
\hline

\multicolumn{4}{|c|}{$\tau = 10$} \\ 
\hline
0.00  & 0.297880  & 0.082568  & 0.215311 \\
4.00  & 0.089687  & 0.077495  & 0.012192 \\
8.00  & -0.675055 & 0.768972  & 1.444027 \\
12.00 & -0.191380 & -0.931082 & 0.739701 \\
16.00 & -0.294592 & -0.921897 & 0.627304 \\
20.00 & 2.404505  & 0.697447  & 1.707057 \\
\hline

\multicolumn{4}{|c|}{$\tau = 16$} \\ 
\hline
0.00  & 0.303434  & 0.089029  & 0.214405 \\
4.00  & -0.118828 & 0.076872  & 0.195700 \\
8.00  & -0.011684 & 0.016508  & 0.028192 \\
12.00 & -0.146955 & -0.323155 & 0.176199 \\
16.00 & 0.758587  & 0.222575  & 0.536012 \\
20.00 & -0.297778 & 0.192386  & 0.490164 \\
\hline

\end{tabular}
 
\end{table}

\begin{table}[ht]
\centering
\caption{Effect of the parameters $\beta$, $\tau$, and $\alpha$ on the solution.}
\label{tab:remarks}
\renewcommand{\arraystretch}{1.4}
\begin{tabular}{|p{2.2cm}|p{11.5cm}|}
\hline
\textbf{Parameter} & \textbf{Observation} \\
\hline

$\beta$ &
Positive and negative values of $\beta$ behave differently only when the delay is small $(\tau=4)$; negative $\beta$ produces much larger numerical values. For a medium delay $(\tau=10)$, both give almost the same results. For a large delay $(\tau=16)$, changing the sign of $\beta$ mainly changes the sign of the solution, while the magnitude remains almost the same. \\
\hline

$\tau$ &
As $\tau$ increases from $4$ to $10$ to $16$, the oscillations and growth in the numerical values reduce significantly. At $\tau=4$, the values are large and change rapidly. At $\tau=10$, they become moderate. At $\tau=16$, the values are much smaller and more stable. Thus, increasing $\tau$ makes the solution calmer and less sensitive. \\
\hline

$\alpha$ &
Although larger values of $\alpha$ produce a smoother solution, a small value $\alpha=0.1$ is chosen to better capture the delay-induced oscillations and fine details of the solution. Increasing $\alpha$ reduces these effects and makes the solution overly smooth. Therefore, $\alpha=0.1$ provides a clearer representation of the actual behaviour of the solution, especially the effects caused by the delay term. \\
\hline

\end{tabular}
\end{table}

\subsection{Solution via Fourier transform}
In this section, we present a method to obtain the solution of the delay differential equation (DDE) using the Offset Linear Canonical Transform (OLCT), based on its connection with the Fourier Transform (FT).
Delay differential equation:
	\begin{eqnarray*}
		\frac{d}{dt}{\hat{f}}(t)+\alpha t{\hat{f}}(t)&=& \beta {\hat{f}}(t-\tau)
			\end{eqnarray*}

			Applying FT on both sides we get,
			\begin{eqnarray*}
				&&\mathcal{F}[\frac{d}{dt}{\hat{f}}(t)]u +\alpha\mathcal{F}[ t{\hat{f}}(t)] u= \beta \mathcal{F}[{\hat{f}}(t-\tau)]u
				\end{eqnarray*}
				Using Properties of FT we can write
					
			\begin{eqnarray}
				iu\,\mathcal{F}[\hat{f}](u)+i\alpha \frac{d}{du}\mathcal{F}[\hat{f}](u)
				&=&\beta \mathcal{F}[\hat{f}](u)e^{-iu\tau}
			\end{eqnarray}
			
			Replacing $u$ by $\frac{u}{b}$ we get
			\begin{eqnarray*}
				i\frac{u}{b}\mathcal{F}[\hat{f}]\left(\frac{u}{b}\right)
				+ i\alpha b \frac{d}{du}\mathcal{F}[\hat{f}]\left(\frac{u}{b}\right)
				= \beta \mathcal{F}[\hat{f}]\left(\frac{u}{b}\right)e^{-i\frac{u}{b}\tau}
			\end{eqnarray*}
			
			Substitute$f(t)e^{\frac{i}{2b}(a t^2+2tu_0)}$ in place of $\hat{f}(t)$
	Then
		\begin{eqnarray*}
		i\frac{u}{b}\mathcal{F}[f(t)e^{\frac{i}{2b}(a t^2+2tu_0)} ]\left(\frac{u}{b}\right)
		+ i\alpha b \frac{d}{du}\mathcal{F}[ f(t)e^{\frac{i}{2b}(a t^2+2tu_0)}]\left(\frac{u}{b}\right)
		= \beta \mathcal{F}[ f(t)e^{\frac{i}{2b}(a t^2+2tu_0)}]\left(\frac{u}{b}\right)e^{-i\frac{u}{b}\tau}
	\end{eqnarray*}
		
			Now Using the relation between OLCT and FT \ref{relation between FT and OLCT} we can write,
			\begin{eqnarray*}
				\frac{iu}{b}\sqrt{ib}e^{-E(u)}\mathcal{O}_M(u)+ib\alpha\frac{d}{du}\sqrt{ib}e^{-E(u)}\mathcal{O}_M(u)=\beta e^{-E(u)}\sqrt{ib}\mathcal{O}_M(u)e^{-i\frac{u}{b}\tau}
			\end{eqnarray*} 
			Rearranging the equation we get
			
			\begin{eqnarray*}
				&&ib\alpha\sqrt{ib}\left(\frac{d}{du}e^{-E(u)}\mathcal{O}_M(u)\right)+\frac{iu}{b}\sqrt{ib}e^{-E(u)}\mathcal{O}_M(u)-\beta e^{-E(u)}\sqrt{ib}\mathcal{O}_M(u)e^{-i\frac{u}{b}\tau}=0\\
				&&ib\alpha\left(-e^{-E(u)}(\frac{d}{du}E(u))\mathcal{O}_M(u)+e^{-E(u)}\frac{d}{du}\mathcal{O}_M(u)\right)+	\frac{iu}{b}e^{-E(u)}\mathcal{O}_M(u)-\beta e^{-E(u)}\mathcal{O}_M(u)e^{-i\frac{u}{b}\tau}=0
			\end{eqnarray*}
	Rearranging the above equation we get
	\begin{eqnarray*}
		&&{i\alpha}{b}\frac{d}{du}\mathcal{O}_M(u)+\left({-i\alpha}{b}\left(\frac{d}{du}E(u)\right)+\frac{iu}{b}-\beta e^{-i\frac{u}{b}\tau}\right)\mathcal{O}_M(u)=0\\
		&&{i\alpha}{b}\frac{d}{du}\mathcal{O}_M(u)+\left(-i b\alpha\left(\frac{i}{b}(-(du_0-b\omega_0)+du)\right)+\frac{iu}{b}-\beta e^{-i\frac{u}{b}\tau}\right)\mathcal{O}_M(u)=0\\
		&&\frac{d}{du}\mathcal{O}_M(u)+\left(\left(-\frac{i}{b}(-(du_0-b\omega_0)+du)\right)+	\frac{u}{b^2\alpha}-\beta\frac{1}{i b\alpha} e^{-i\frac{u}{b}\tau}\right)\mathcal{O}_M(u)=0
	\end{eqnarray*}
	This is linear ordinary differential equation.
	
	\begin{eqnarray*}
		\frac{d}{du}\mathcal{O}_M(u)
		+\left(\left(-\frac{i}{b}(-(du_0-b\omega_0)+du)\right)+\frac{u}{b^2\alpha}-\beta\frac{1}{i b\alpha} e^{-i\frac{u}{b}\tau}\right)\mathcal{O}_M(u)=0
	\end{eqnarray*}
	
	So comparing with
	\begin{eqnarray*}
		\frac{dy}{du}+P(u)y=0
	\end{eqnarray*}
	
	we get
	\begin{eqnarray*}
		P(u)=\left(-\frac{i}{b}(-(du_0-b\omega_0)+du)\right)+\frac{u}{b^2\alpha}-\beta\frac{1}{i b\alpha} e^{-i\frac{u}{b}\tau}
	\end{eqnarray*}
	
	Integrating factor is
	\begin{eqnarray*}
		\mu(u)&=&\exp\left(\int P(u)\,du\right)\\
		&=&\exp\Bigg[
		-\frac{i}{b}\left(-(du_0-b\omega_0)u+\frac{d u^2}{2}\right)
		+\frac{u^2}{2 b^2\alpha}
		-\frac{\beta}{\alpha\tau} e^{-i\frac{u}{b}\tau}
		\Bigg]
	\end{eqnarray*}
	
	Hence the solution is
	\begin{eqnarray}
		\mathcal{O}_M(u)
		&=& C\,e^{-\int P(u)\,du}\nonumber\\
		&=& C\,\exp\Bigg[
		\frac{i}{b}\left(-(du_0-b\omega_0)u+\frac{d u^2}{2}\right)
		-\frac{u^2}{2 b^2\alpha}
		+\frac{\beta}{\alpha\tau} e^{-i\frac{u}{b}\tau}
		\Bigg]\ \label{via ft solution}
	\end{eqnarray} 

  \begin{figure}[h]
\centering
\begin{tabular}{cc}
\includegraphics[width=5.5cm]{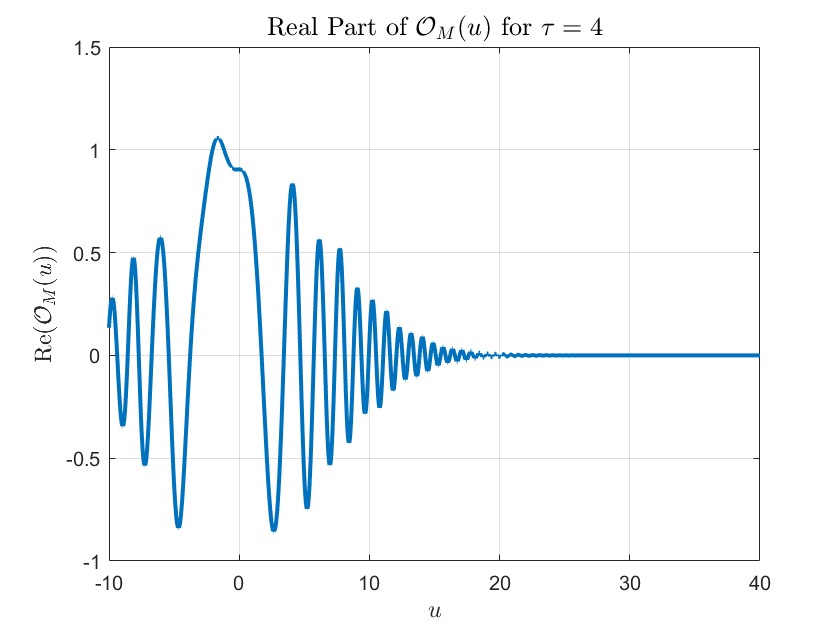} &
\includegraphics[width=5.5cm]{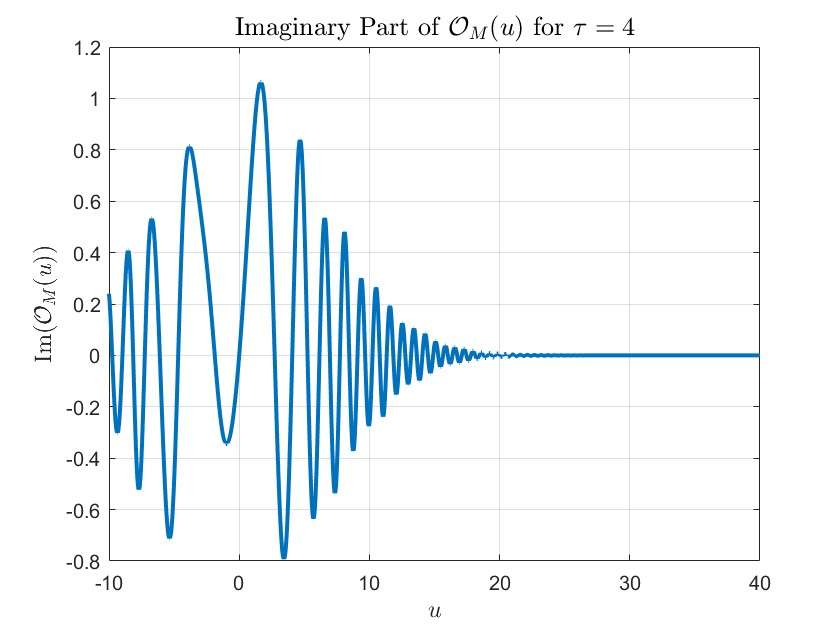} \\

(a) Real part at $\tau=4$, $\beta=-4$, $\alpha=10$ &
(b) Imaginary part at $\tau=4$, $\beta=-4$, $\alpha=10$ \\[0.3cm]

\includegraphics[width=5.5cm]{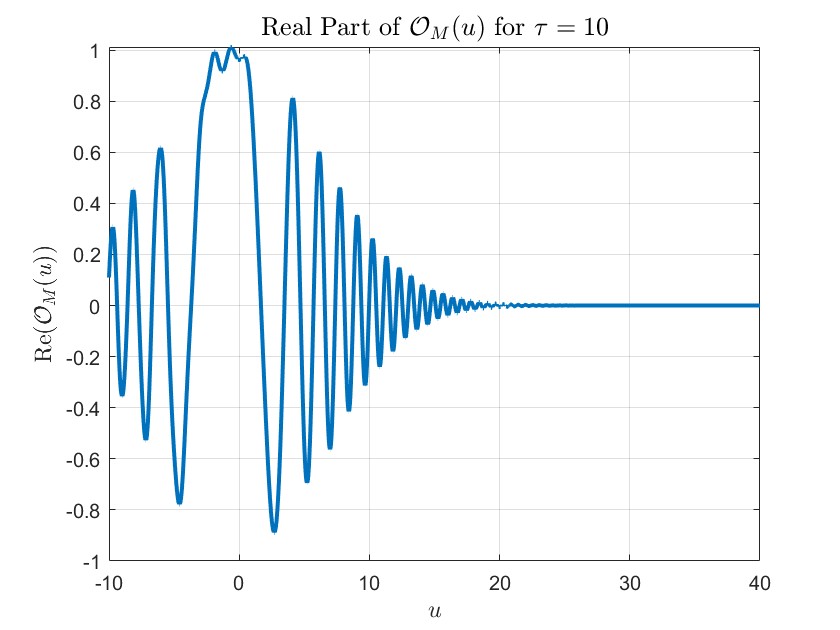} &
\includegraphics[width=5.5cm]{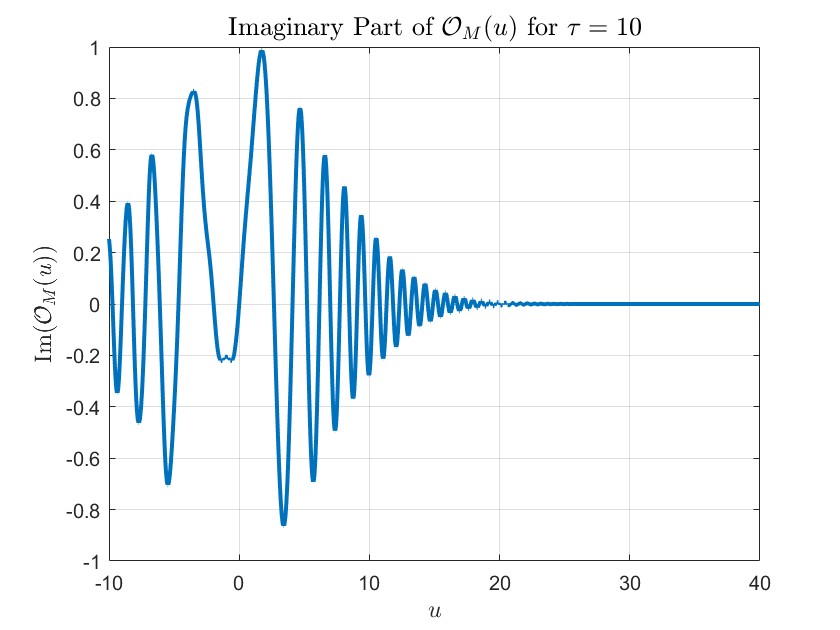} \\

(c) Real part at $\tau=10$, $\beta=-4$, $\alpha=10$ &
(d) Imaginary part at $\tau=10$, $\beta=-4$, $\alpha=10$ \\[0.3cm]

\includegraphics[width=5.5cm]{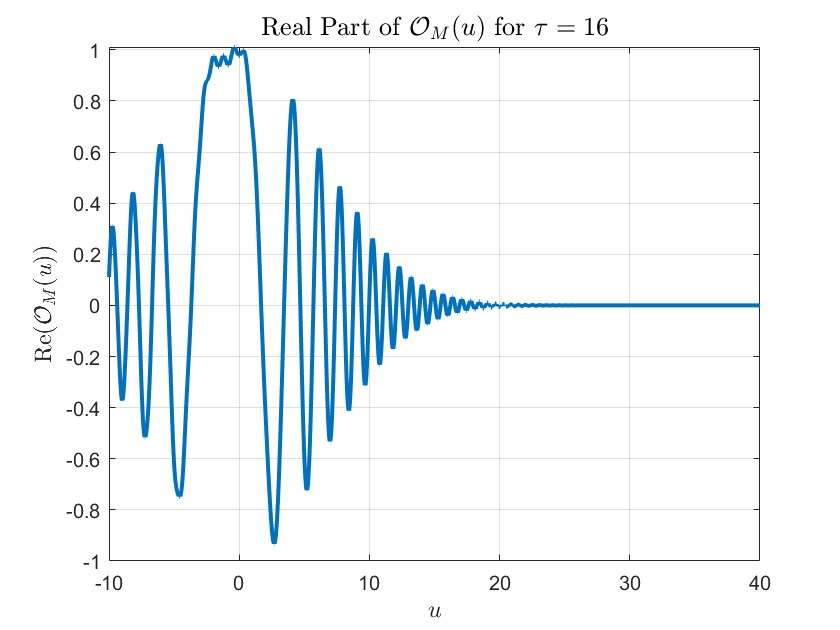} &
\includegraphics[width=5.5cm]{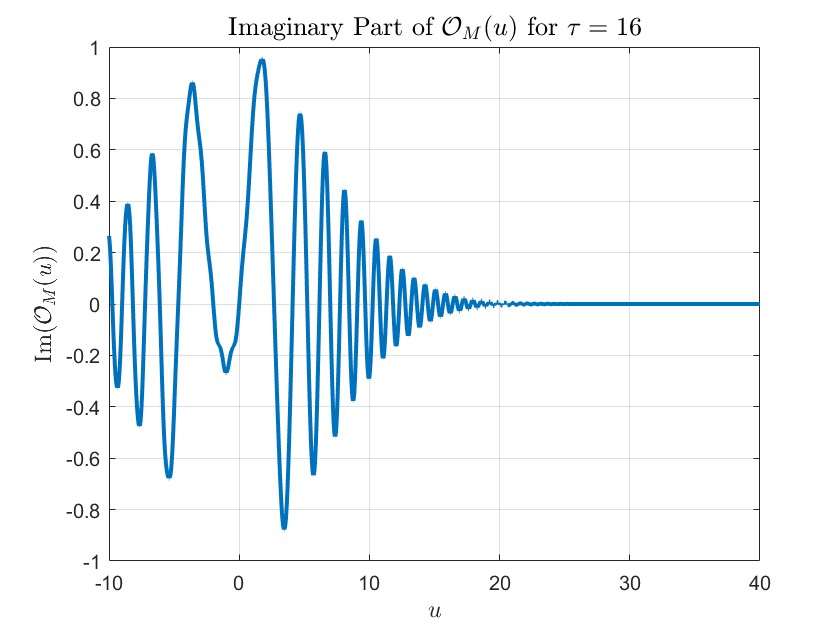} \\

(e) Real part at $\tau=16$, $\beta=-4$, $\alpha=10$ &
(f) Imaginary part at $\tau=16$, $\beta=-4$, $\alpha=10$
\end{tabular}

\caption{Real and imaginary parts of the OLCT-domain solution obtained via the Fourier approach for different values of the delay parameter $\tau$ with $\beta=-4$ and $\alpha=10$ for equation \eqref{via ft solution}.}
\label{fig:via_fourier_olct_negative}
\end{figure}
    
\begin{table}[ht]
\centering
\caption{Numerical values for different values of $\tau$(beta negative) for equation \eqref{via ft solution}.}
\label{tab:tau_olct}
\begin{tabular}{|c|c|c|c|}
\hline
$u$ & Real Part & Imag Part & Difference \\
\hline

\multicolumn{4}{|c|}{$\tau = 4$} \\
\hline
4.00  & 0.816669  & -0.152197 & 0.968867 \\
8.00  & 0.214703  & 0.445444  & -0.230741 \\
12.00 & -0.076243 & -0.138882 & 0.062639 \\
16.00 & -0.036741 & 0.007505  & -0.044246 \\
20.00 & -0.007152 & -0.000853 & -0.006298 \\
\hline

\multicolumn{4}{|c|}{$\tau = 10$} \\
\hline
4.00  & 0.781094  & -0.196676 & 0.977770 \\
8.00  & 0.175682  & 0.426728  & -0.251046 \\
12.00 & -0.070598 & -0.156534 & 0.085936 \\
16.00 & -0.039157 & 0.011958  & -0.051115 \\
20.00 & -0.006508 & -0.000156 & -0.006351 \\
\hline

\multicolumn{4}{|c|}{$\tau = 16$} \\
\hline
4.00  & 0.772911  & -0.213410 & 0.986321 \\
8.00  & 0.172185  & 0.410282  & -0.238097 \\
12.00 & -0.062654 & -0.153772 & 0.091118 \\
16.00 & -0.040299 & 0.009802  & -0.050102 \\
20.00 & -0.006896 & -0.000343 & -0.006552 \\
\hline

\end{tabular}
\end{table}

\noindent\textbf{Observation.}
From Table~\ref{tab:tau_olct}, it is observed that both the real and imaginary parts exhibit oscillatory behaviour as the delay parameter $\tau$ varies. The quantity $\mathrm{Re}(\mathcal{O}_M(u))-\mathrm{Im}(\mathcal{O}_M(u))$ changes sign for different values of $u$, indicating alternating dominance between the real and imaginary components. Moreover, the magnitude of the solution decreases as $u$ increases, demonstrating a damped oscillatory behaviour. The influence of the delay parameter is more noticeable for smaller values of $u$, whereas for larger values of $u$ the solution values become very small and tend towards zero. 

\begin{figure}[h]
\centering
\begin{tabular}{cc}
\includegraphics[width=5.5cm]{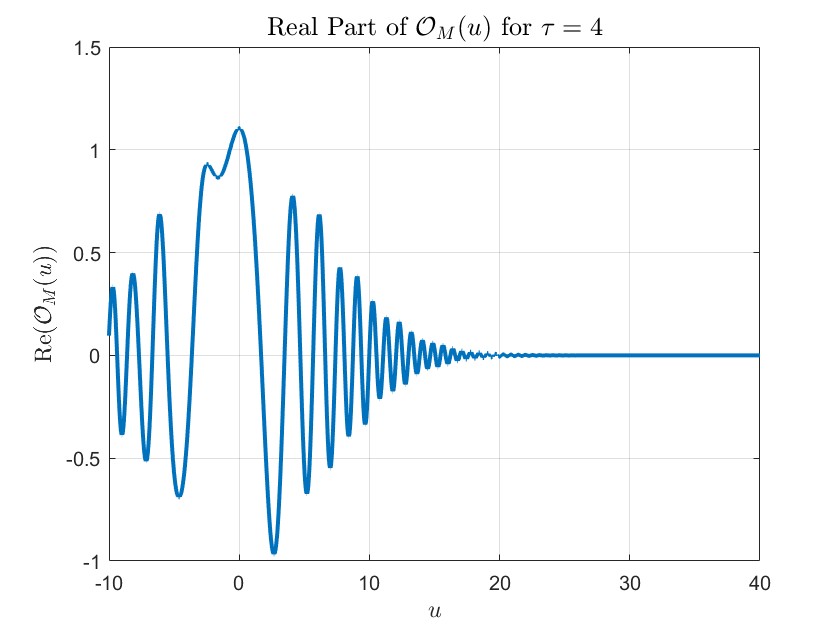} &
\includegraphics[width=5.5cm]{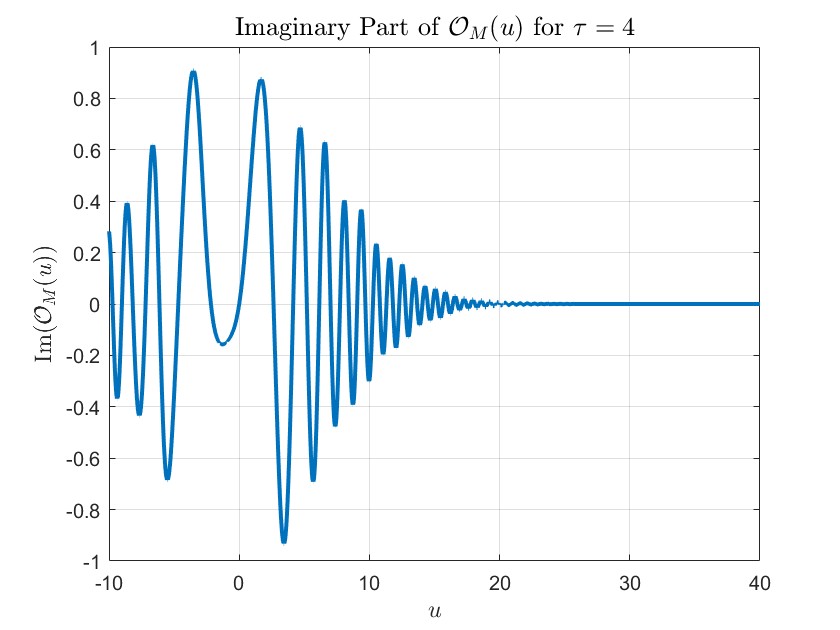} \\

(a) Real part at $\tau=4$, $\beta=4$, $\alpha=10$ &
(b) Imaginary part at $\tau=4$, $\beta=4$, $\alpha=10$ \\[0.3cm]

\includegraphics[width=5.5cm]{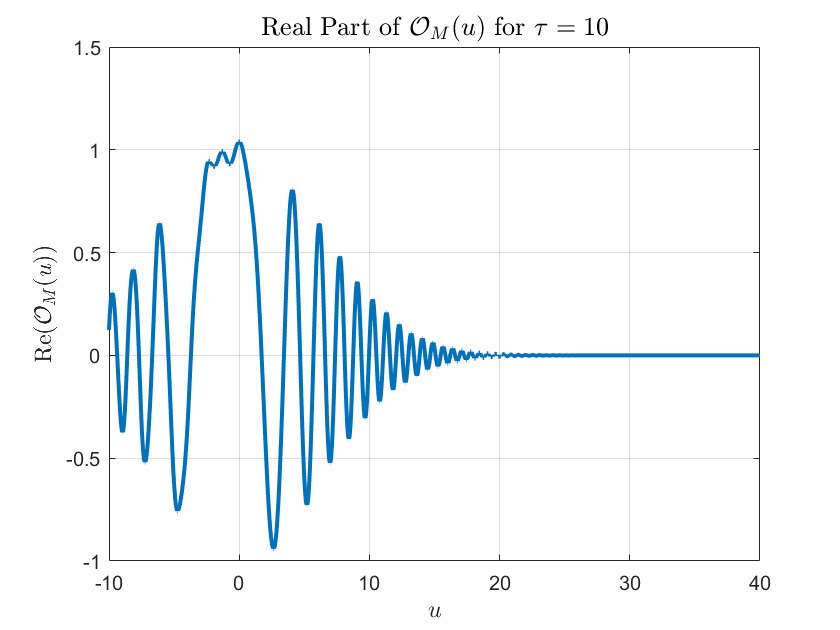} &
\includegraphics[width=5.5cm]{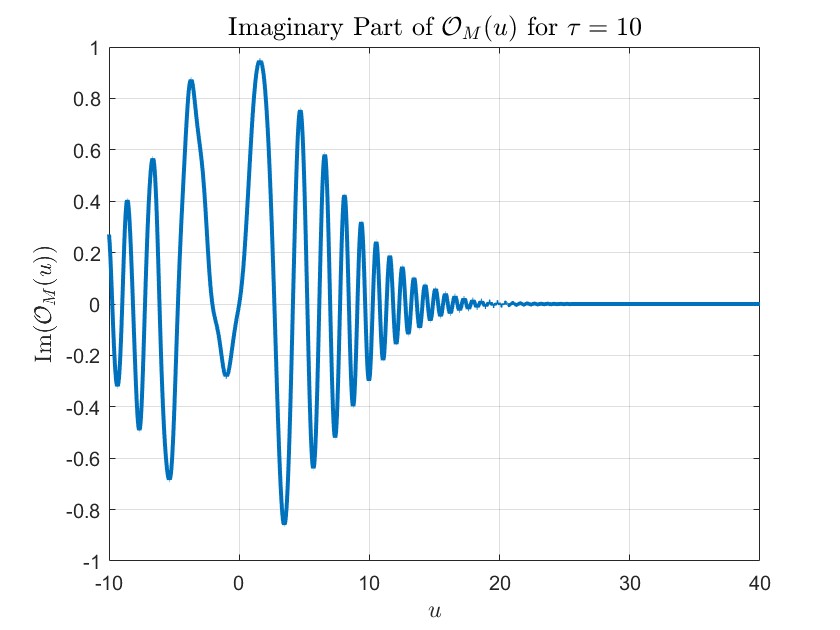} \\

(c) Real part at $\tau=10$, $\beta=4$, $\alpha=10$ &
(d) Imaginary part at $\tau=10$, $\beta=4$, $\alpha=10$ \\[0.3cm]

\includegraphics[width=5.5cm]{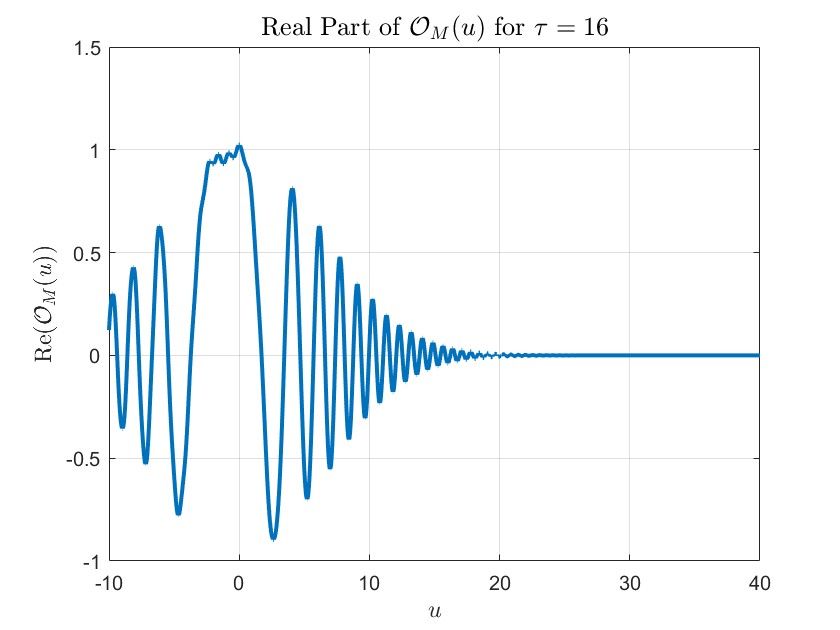} &
\includegraphics[width=5.5cm]{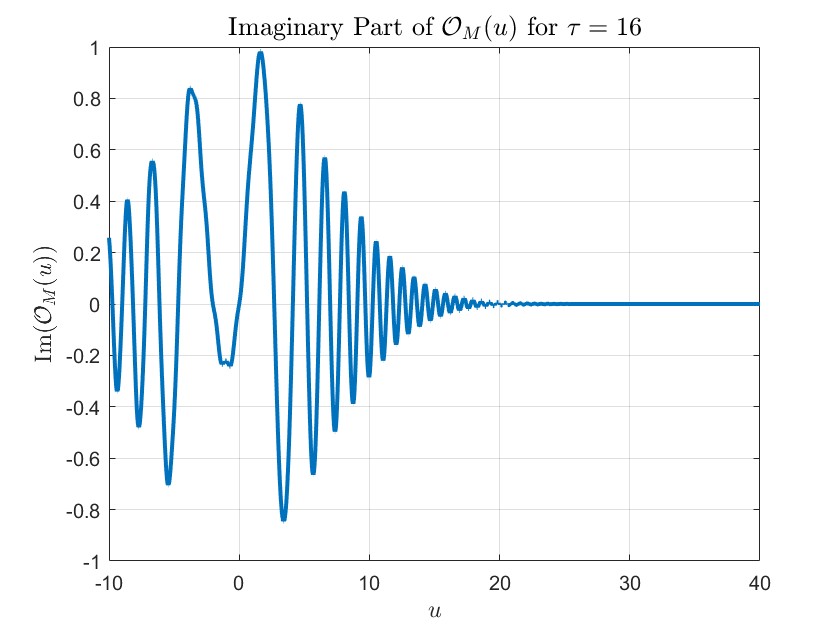} \\

(e) Real part at $\tau=16$, $\beta=4$, $\alpha=10$ &
(f) Imaginary part at $\tau=16$, $\beta=4$, $\alpha=10$
\end{tabular}

\caption{Real and imaginary parts of the OLCT-domain solution obtained via the Fourier approach for different values of the delay parameter $\tau$ with $\beta=4$ and $\alpha=10$ for equation \eqref{via ft solution}.}
\label{fig:via_fourier_olct_positive}
\end{figure}

\begin{table}[ht]
\centering
\caption{Numerical values for different values of $\tau$ for equation \eqref{via ft solution}.}
\label{tab:tau_values}
\begin{tabular}{|c|c|c|c|}
\hline
$u$ & Real Part & Imag Part & Difference \\
\hline

\multicolumn{4}{|c|}{$\tau = 4$} \\
\hline
4.00  & 0.748707 & -0.300886 & 1.049593 \\
8.00  & 0.155819 &  0.377393 & -0.221574 \\
12.00 & -0.054405 & -0.163655 & 0.109250 \\
16.00 & -0.042172 &  0.013592 & -0.055764 \\
20.00 & -0.006300 &  0.000191 & -0.006491 \\
\hline

\multicolumn{4}{|c|}{$\tau = 10$} \\
\hline
4.00  & 0.790036 & -0.261549 & 1.051585 \\
8.00  & 0.190359 &  0.393916 & -0.203557 \\
12.00 & -0.061863 & -0.146602 & 0.084739 \\
16.00 & -0.039632 &  0.008733 & -0.048364 \\
20.00 & -0.006960 & -0.000450 & -0.006510 \\
\hline

\multicolumn{4}{|c|}{$\tau = 16$} \\
\hline
4.00  & 0.799392 & -0.244630 & 1.044022 \\
8.00  & 0.194646 &  0.409883 & -0.215236 \\
12.00 & -0.069508 & -0.149154 & 0.079646 \\
16.00 & -0.038561 &  0.010866 & -0.049426 \\
20.00 & -0.006571 & -0.000255 & -0.006316 \\
\hline

\end{tabular}
\end{table} 

Now	from the inverse of OLCT we Know
	
 \begin{eqnarray*}
	f(t)&=&\frac{1}{\sqrt{2\pi (-i) b}}\int_{\mathbb{R}}e^{\frac{-i}{2b}(a t^2 + 2 t (u_0-u)-2u(du_0-b\omega_0)+ du^2+ du_0^2)}\mathcal{O}_M(u)\ du\\
	&=&\frac{1}{\sqrt{2\pi (-i) b}}\int_{\mathbb{R}}e^{\frac{-i}{2b}(a t^2 + 2 t (u_0-u)-2u(du_0-b\omega_0)+ du^2+ du_0^2)}\\&\times&  C\,\exp\Bigg[
	\frac{i}{b}\left(-(du_0-b\omega_0)u+\frac{d u^2}{2}\right)
	-\frac{u^2}{2 b^2\alpha}
	+\frac{\beta}{\alpha\tau} e^{-i\frac{u}{b}\tau}
	\Bigg]
	\end{eqnarray*}
	Combine exponential terms
\begin{eqnarray}
f(t)&=&\frac{C}{\sqrt{2\pi (-i) b}}e^{\frac{-i}{2b}(a t^2 + 2t u_0 + du_0^2)}\int_{\mathbb{R}}\exp\Bigg[-\frac{u^2}{2 b^2\alpha}+ \frac{i t}{b}u	+\frac{\beta}{\alpha\tau} e^{-i\frac{u}{b}\tau}	\Bigg] du\nonumber\\
	&=&\frac{C}{\sqrt{2\pi (-i) b}}e^{\frac{-i}{2b}(a t^2 + 2t u_0 + d u_0^2)}
\int_{\mathbb{R}}e^{\left(-\frac{u^2}{2 b^2\alpha}	+ \frac{i t}{b}u\right)}
e^{\left(\frac{\beta}{\alpha\tau} e^{-i\frac{u}{b}\tau}\right)} du\nonumber\\
&=&\frac{C}{\sqrt{2\pi (-i) b}}e^{\frac{-i}{2b}(a t^2 + 2t u_0 + d u_0^2)}
\int_{\mathbb{R}}e^{\left(-\frac{u^2}{2 b^2\alpha}+ \frac{\beta}{\alpha\tau}\cos\left(\frac{u\tau}{b}\right)
	\right)}
e^{i\left(\frac{t}{b}u- \frac{\beta}{\alpha\tau}\sin\left(\frac{u\tau}{b}\right)
	\right)} du\nonumber\nonumber\\
	&=&\frac{C}{\sqrt{2\pi b}}	\int_{\mathbb{R}}	e^{\left(-\frac{u^2}{2b^2\alpha}	+\frac{\beta}{\alpha\tau}\cos\left(\frac{u\tau}{b}\right)
		\right)}\times\Bigg[\cos\left(\frac{\pi}{4}
	-\frac{1}{2b}(a t^2 + 2t u_0 + d u_0^2)
	+\frac{t}{b}u-\frac{\beta}{\alpha\tau}\sin\left(\frac{u\tau}{b}\right)
	\right)\nonumber\\
	&&+ i \sin\left(
	\frac{\pi}{4}-\frac{1}{2b}(a t^2 + 2t u_0 + d u_0^2)
	+\frac{t}{b}u-\frac{\beta}{\alpha\tau}\sin\left(\frac{u\tau}{b}\right)\right)
	\Bigg] du\label{OLCT solution via fourier}
	\end{eqnarray}
	
	For large values of $\tau$, the solution becomes increasingly smooth and Gaussian-like in nature.
    
\begin{eqnarray*}
	\lim_{\tau \to \infty} f(t)
	&=&\frac{C}{\sqrt{2\pi b}}
	\int_{\mathbb{R}}
	e^{-\frac{u^2}{2 b^2\alpha}}
	e^{i\left(
		\frac{\pi}{4}
		-\frac{1}{2b}(a t^2 + 2t u_0 + d u_0^2)
		+\frac{t}{b}u
		\right)} du\\
&=&\frac{C}{\sqrt{2\pi b}}
	e^{i\left(
		\frac{\pi}{4}
		-\frac{1}{2b}(a t^2 + 2t u_0 + d u_0^2)
		\right)}
	\int_{\mathbb{R}}
	e^{-\frac{u^2}{2 b^2\alpha}}
	e^{i\frac{t}{b}u} du
\end{eqnarray*}
\begin{eqnarray*}
	\lim_{\tau \to \infty} f(t)
	&=&C\sqrt{\alpha b}\
	e^{\left[
	i\left(\frac{\pi}{4}
	-\frac{1}{2b}(a t^2 + 2t u_0 + d u_0^2)\right)
	\right]}
	e^{\left(-\frac{\alpha t^2}{2}\right)}
\end{eqnarray*}

\begin{figure}[h]
\centering
\begin{tabular}{cc}
\includegraphics[width=6.5cm]{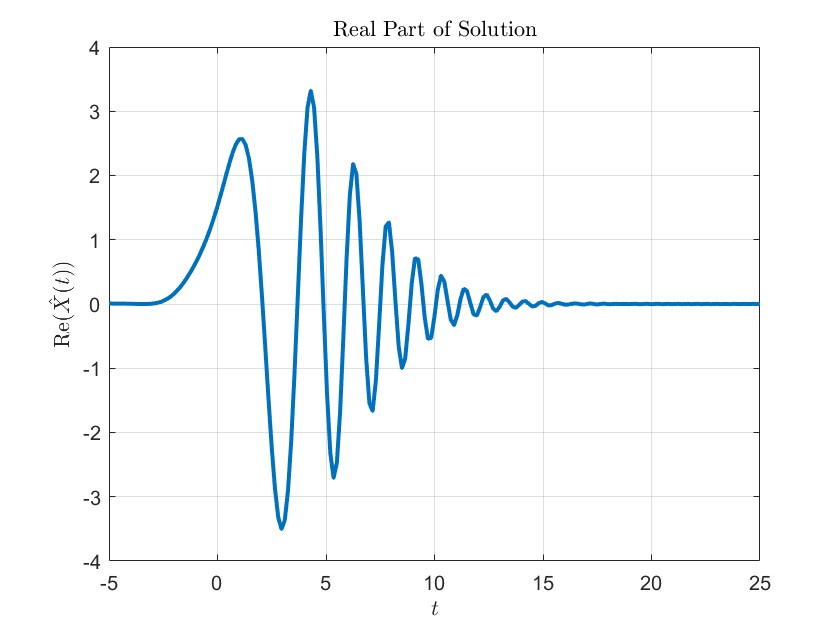} &
\includegraphics[width=6.5cm]{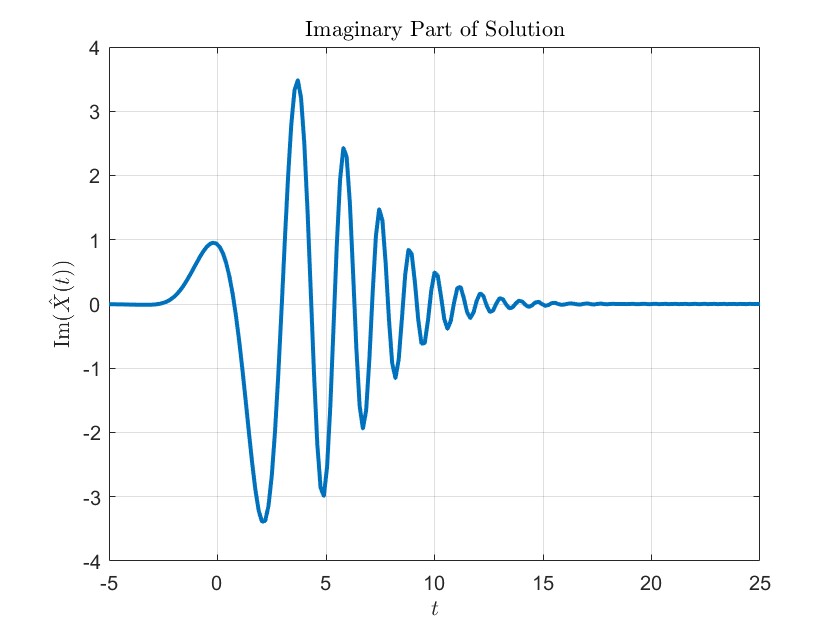} \\

(a) Real part at $\tau=2$, $\beta=4$, $\alpha=1$ &
(b) Imaginary part at $\tau=2$, $\beta=4$, $\alpha=1$ \\[0.3cm]

\includegraphics[width=6.5cm]{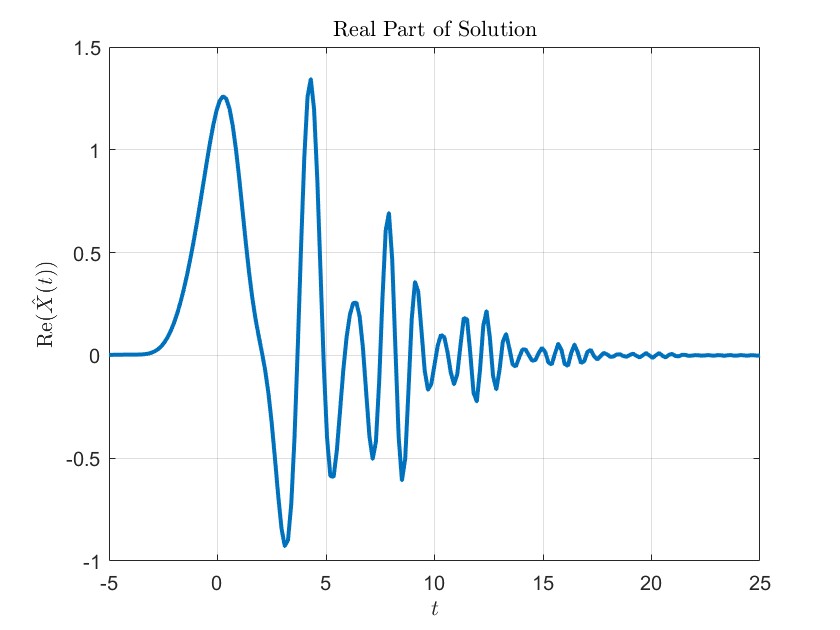} &
\includegraphics[width=6.5cm]{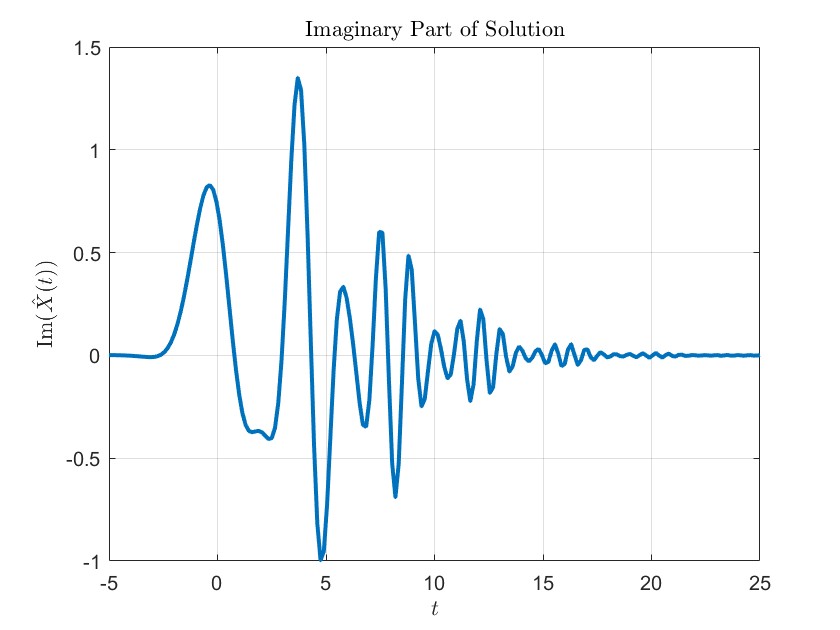} \\

(c) Real part at $\tau=4$, $\beta=4$, $\alpha=1$ &
(d) Imaginary part at $\tau=4$, $\beta=4$, $\alpha=1$ \\[0.3cm]

\includegraphics[width=6.5cm]{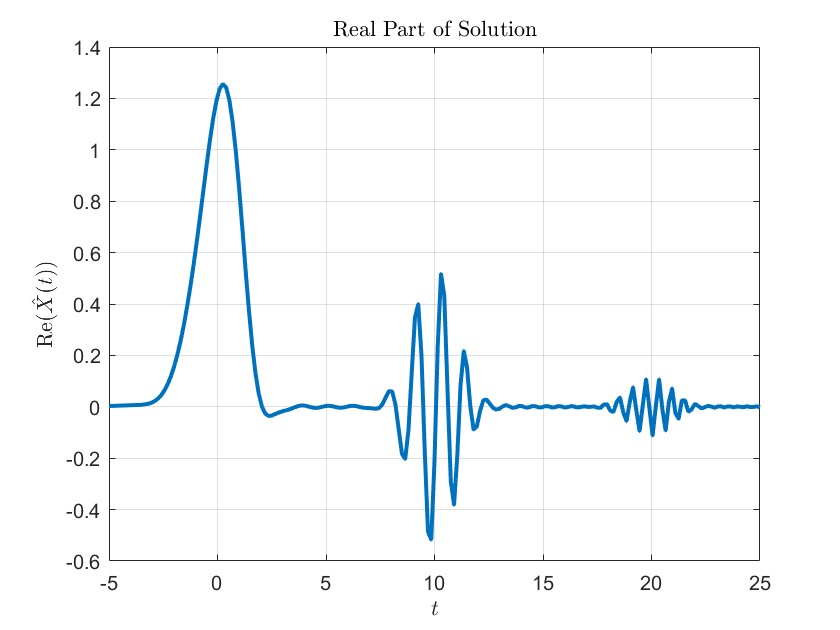} &
\includegraphics[width=6.5cm]{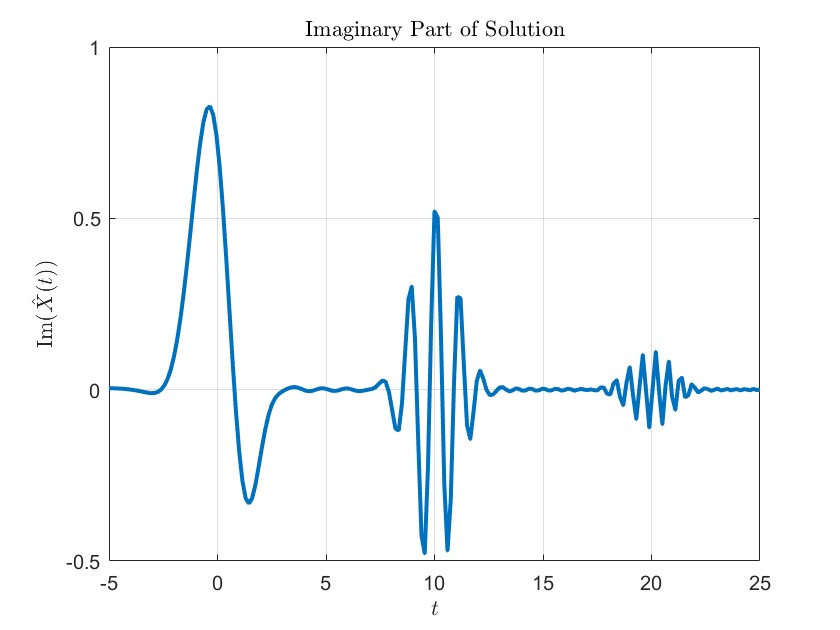} \\

(e) Real part at $\tau=10$, $\beta=4$, $\alpha=1$ &
(f) Imaginary part at $\tau=10$, $\beta=4$, $\alpha=1$
\end{tabular}

\caption{Real and imaginary parts of the solution obtained via the Fourier-transform approach for $\beta=4$ and $\alpha=1$ corresponding to different values of the delay parameter $\tau$ for equation \eqref{OLCT solution via fourier}.}
\label{fig:via_ft_positive}
\end{figure}

\begin{table}[h]
\centering
\caption{Numerical values for different values of $\tau$ (Negative $\beta$) for equation \eqref{OLCT solution via fourier}}
\renewcommand{\arraystretch}{1.4}
\begin{tabular}{|c|c|c|c|}
\hline
$t$ & Real Part & Imag Part & Difference \\
\hline

\multicolumn{4}{|c|}{$\tau = 2$} \\
\hline
0.05  & 0.865723  & 0.481907  & 0.383816 \\
9.95  & 0.135768  & -0.175235 & 0.311004 \\
20.00 & -0.000005 & 0.000563  & 0.000568 \\
30.05 & -0.000349 & 0.000266  & 0.000615 \\
39.95 & 0.000286  & 0.000177  & 0.000110 \\
\hline

\multicolumn{4}{|c|}{$\tau = 4$} \\
\hline
0.05  & 1.227789  & 0.680402  & 0.547387 \\
9.95  & -0.043012 & 0.055925  & 0.098938 \\
20.00 & 0.013724  & 0.005786  & 0.007938 \\
30.05 & 0.000959  & 0.001947  & 0.000988 \\
39.95 & 0.001044  & -0.001166 & 0.002210 \\
\hline

\multicolumn{4}{|c|}{$\tau = 10$} \\
\hline
0.05  & 1.224670  & 0.686965  & 0.537705 \\
9.95  & 0.340811  & -0.447798 & 0.788609 \\
20.00 & -0.098934 & -0.053213 & 0.045722 \\
30.05 & -0.000924 & 0.015161  & 0.016084 \\
39.95 & -0.001166 & 0.000500  & 0.001666 \\
\hline

\end{tabular}\label{table via ft negative}
\end{table}

\begin{figure}[h]
\centering
\begin{tabular}{cc}
\includegraphics[width=5.5cm]{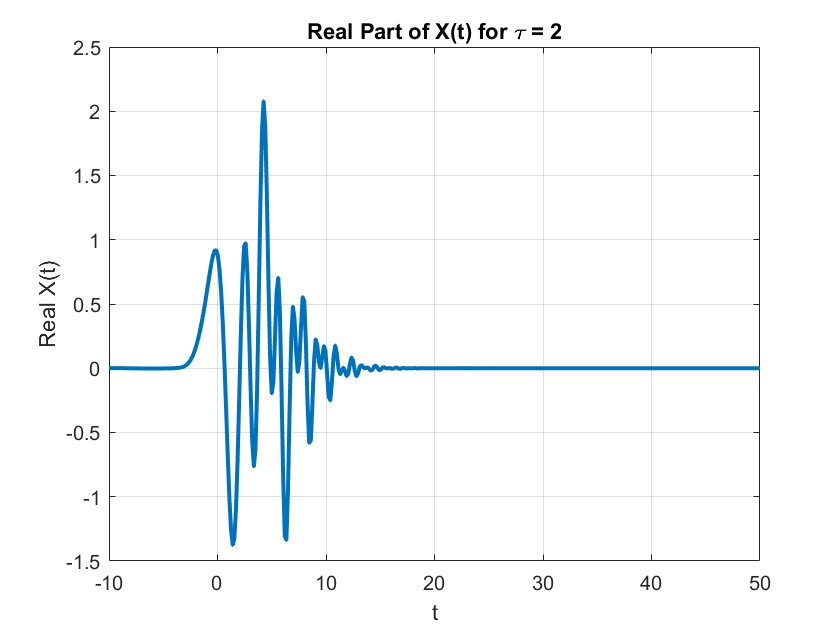} &
\includegraphics[width=5.5cm]{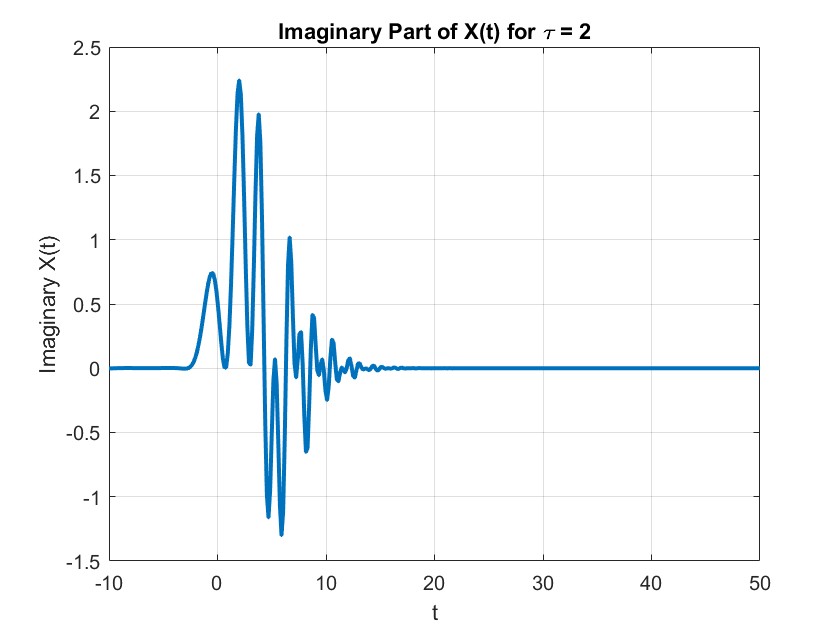} \\

(a) Real part at $\tau=2$, $\beta=-4$, $\alpha=1$ &
(b) Imaginary part at $\tau=2$, $\beta=-4$, $\alpha=1$ \\[0.3cm]

\includegraphics[width=5.5cm]{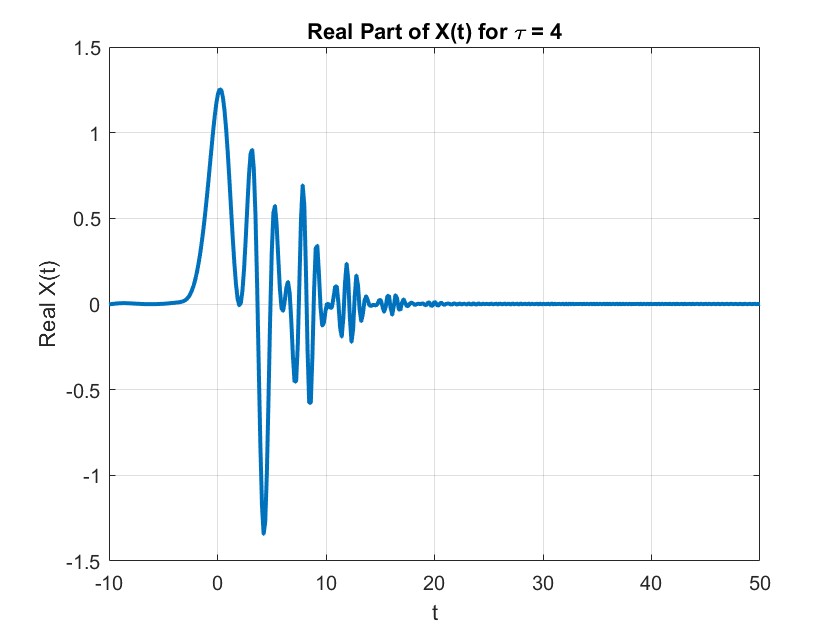} &
\includegraphics[width=5.5cm]{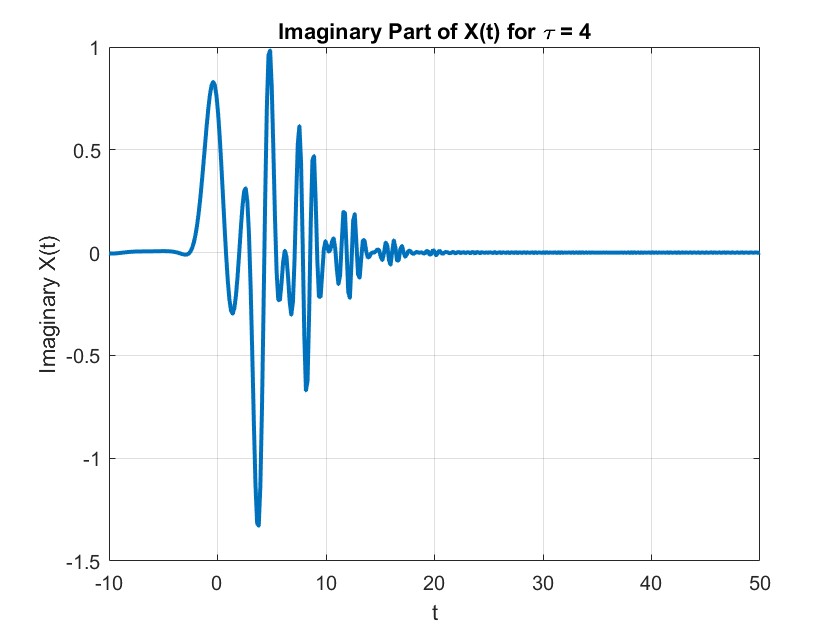} \\

(c) Real part at $\tau=4$, $\beta=-4$, $\alpha=1$ &
(d) Imaginary part at $\tau=4$, $\beta=-4$, $\alpha=1$ \\[0.3cm]

\includegraphics[width=5.5cm]{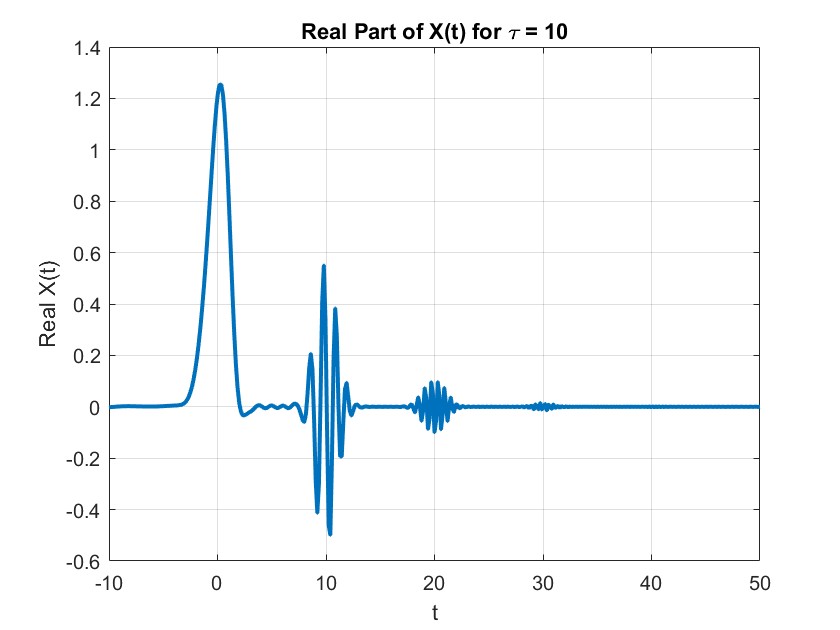} &
\includegraphics[width=5.5cm]{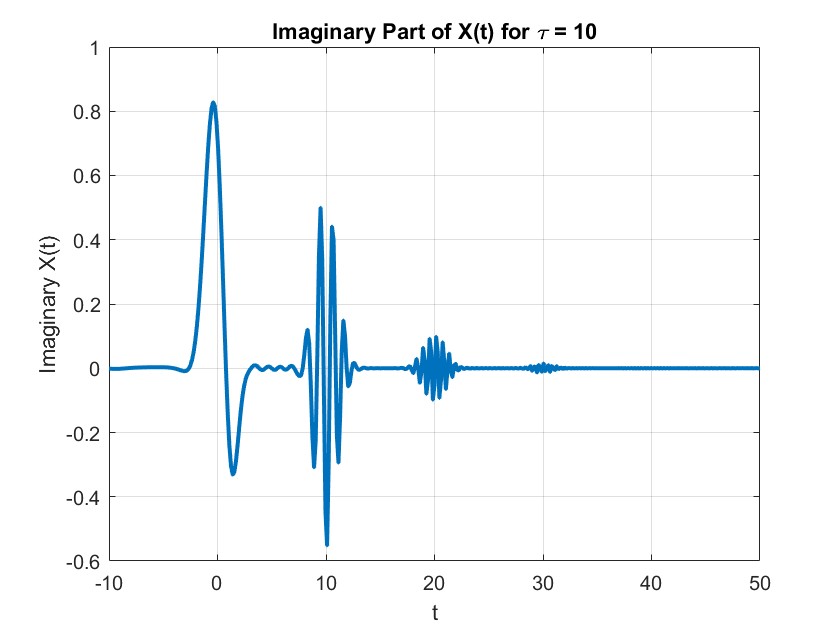} \\

(e) Real part at $\tau=10$, $\beta=-4$, $\alpha=1$ &
(f) Imaginary part at $\tau=10$, $\beta=-4$, $\alpha=1$
\end{tabular}

\caption{Real and imaginary parts of the solution obtained via the Fourier-transform approach for $\beta=-4$ and $\alpha=1$ corresponding to different values of the delay parameter $\tau$ for equation \eqref{OLCT solution via fourier}.}
\label{fig:via_ft_negative}
\end{figure}

\begin{table}[h]
\centering
\caption{Numerical values for different values of $\tau$ (Positive $\beta$) for equation \eqref{OLCT solution via fourier}}

\renewcommand{\arraystretch}{1.4}
\begin{tabular}{|c|c|c|c|}
\hline
$t$ & Real Part & Imag Part & Difference \\
\hline

\multicolumn{4}{|c|}{$\tau = 2$} \\
\hline
0.05  & 1.581673  & 0.910132  & 0.671541 \\
9.95  & -0.323017 & 0.431124  & 0.754141 \\
20.00 & -0.002814 & -0.001035 & 0.001778 \\
30.05 & -0.000059 & -0.001511 & 0.001452 \\
39.95 & -0.001060 & 0.000375  & 0.001434 \\
\hline

\multicolumn{4}{|c|}{$\tau = 4$} \\
\hline
0.05  & 1.224453  & 0.690837  & 0.533616 \\
9.95  & -0.080396 & 0.106587  & 0.186983 \\
20.00 & -0.010332 & -0.005945 & 0.004387 \\
30.05 & 0.000314  & -0.000095 & 0.000409 \\
39.95 & -0.000116 & -0.000215 & 0.000099 \\
\hline

\multicolumn{4}{|c|}{$\tau = 10$} \\
\hline
0.05  & 1.224670  & 0.684998  & 0.539672 \\
9.95  & -0.342383 & 0.443474  & 0.785857 \\
20.00 & -0.097453 & -0.054407 & 0.043046 \\
30.05 & 0.002763  & -0.014315 & 0.017077 \\
39.95 & -0.001172 & -0.000211 & 0.000961 \\
\hline

\end{tabular} \label{table via ft positive}
\end{table}

\begin{table}[ht]
\centering
\caption{Summary of the effects of the parameters $\tau$, $\beta$, and $\alpha$ obtained from Figures \ref{fig:via_ft_negative} ,\ref{fig:via_ft_positive} and Tables \ref{table via ft negative}, \ref{table via ft positive}.}
\label{tab:parameter_effects}
\renewcommand{\arraystretch}{1.4}
\begin{tabular}{|p{2.5cm}|p{11.5cm}|}
\hline
\textbf{Parameter} & \textbf{Observation} \\
\hline

$\tau$ &
From both the positive and negative $\beta$ figures (\ref{fig:via_ft_negative} and \ref{fig:via_ft_positive}) and the numerical Tables~\ref{table via ft negative} and \ref{table via ft positive}, it is observed that as the delay parameter $\tau$ increases, the oscillations become more delayed and gradually decay. For larger values of $\tau$, the solution exhibits smaller amplitudes and a decaying behaviour. \\
\hline

$\beta$ &
The sign of $\beta$ mainly affects the solution for small values of $\tau$. For larger values of $\tau$, the solutions corresponding to positive and negative $\beta$ become very similar, indicating that the influence of the sign of $\beta$ decreases as the delay increases. \\
\hline

$\alpha$ &
The parameter $\alpha$ controls the oscillatory nature of the solution. Larger values of $\alpha$ increase the oscillations, whereas smaller values produce smoother behaviour. \\
\hline

\end{tabular}
\end{table}

\FloatBarrier

\section{Comparison analysis}

In the previous sections, the solution was obtained using two different methods: the direct OLCT method and the Fourier-transform-based method. In this section, we compute the error between the corresponding solutions to compare the two approaches. The error analysis helps to verify the accuracy and validity of the proposed OLCT method.
\begin{table}[ht]
\centering
\caption{Comparison of direct OLCT-domain solution and Fourier-based solution for different values of $\tau$(beta negative) for equation \eqref{olct domain solution} and \eqref{via ft solution}.}
\label{tab:comparison}
\renewcommand{\arraystretch}{1.4}

\begin{tabular}{|c|c|c|c|c|}
\hline
$u$ & Direct Re & Fourier Re & Direct Im & Fourier Im \\
\hline
\multicolumn{5}{|c|}{$\tau=4$} \\
\hline
4  & 0.82960 & 0.81667 & -0.17934 & -0.15220 \\
8  & 0.25914 & 0.21470 &  0.39681 &  0.44544 \\
12 & -0.08248 & -0.07624 & -0.07287 & -0.13888 \\
16 & -0.02453 & -0.03674 &  0.00150 &  0.00750 \\
20 & -0.00654 & -0.00715 & -0.00028 & -0.00085 \\
\hline
\multicolumn{5}{|c|}{$\tau=10$} \\
\hline
4  & 0.73454 & 0.78109 & -0.22040 & -0.19668 \\
8  & 0.20498 & 0.17568 &  0.34810 &  0.42673 \\
12 & -0.09161 & -0.07060 & -0.08632 & -0.15653 \\
16 & -0.00393 & -0.03916 &  0.01122 &  0.01196 \\
20 & 0.00152 & -0.00651 & -0.00540 & -0.00016 \\
\hline
\multicolumn{5}{|c|}{$\tau=16$} \\
\hline
4  & 0.70974 & 0.77291 & -0.22203 & -0.21341 \\
8  & 0.17908 & 0.17219 &  0.32446 &  0.41028 \\
12 & -0.07459 & -0.06265 & -0.09071 & -0.15377 \\
16 & -0.00227 & -0.04030 &  0.01830 &  0.00980 \\
20 & 0.01352 & -0.00690 &  0.00070 & -0.00034 \\
\hline
\end{tabular}\label{table camparision of two solution negative beta}
\end{table}

 \begin{figure}[h]
\centering
\begin{tabular}{cc}
\includegraphics[width=8.5cm]{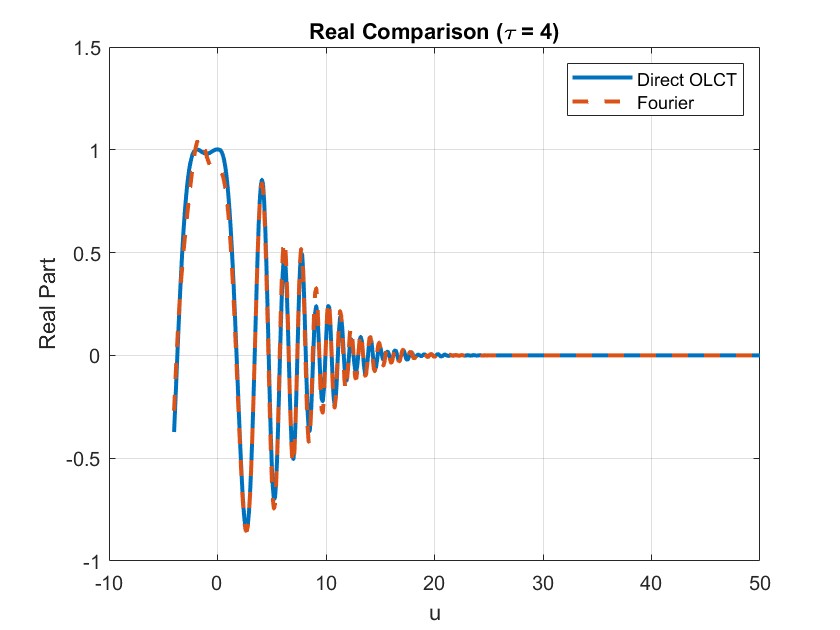} &
\includegraphics[width=8.5cm]{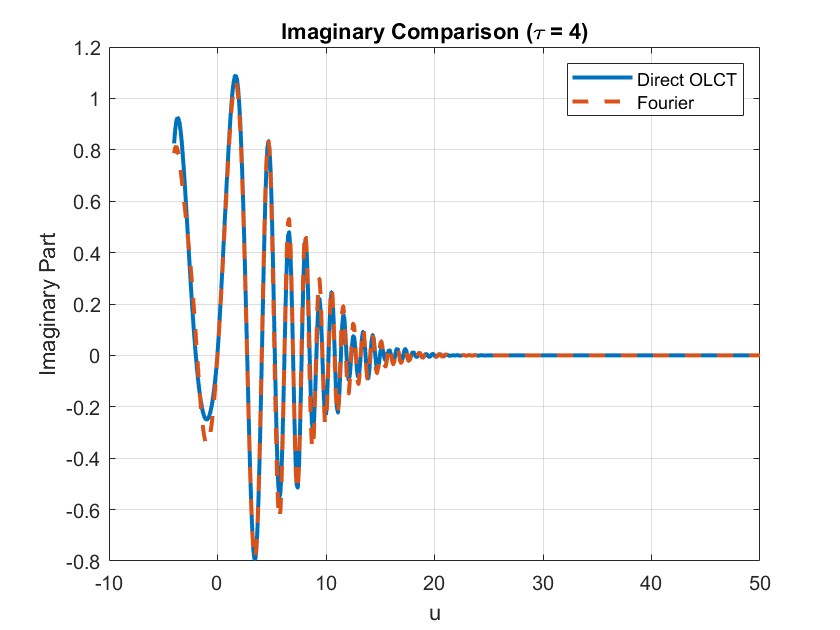} \\
(a) Real part at $\tau=4$ &
(b) Imaginary part at $\tau=4$ \\[0.3cm]

\includegraphics[width=8.5cm]{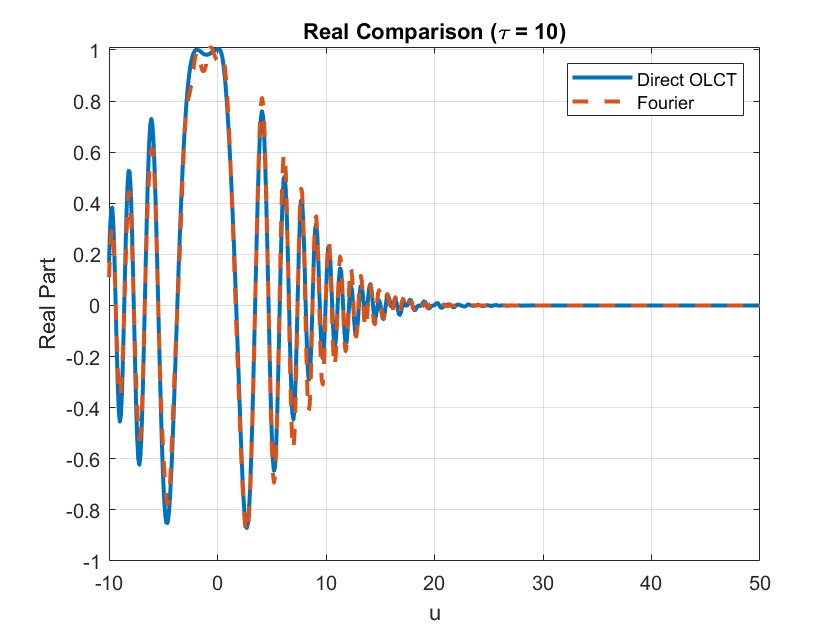} &
\includegraphics[width=8.5cm]{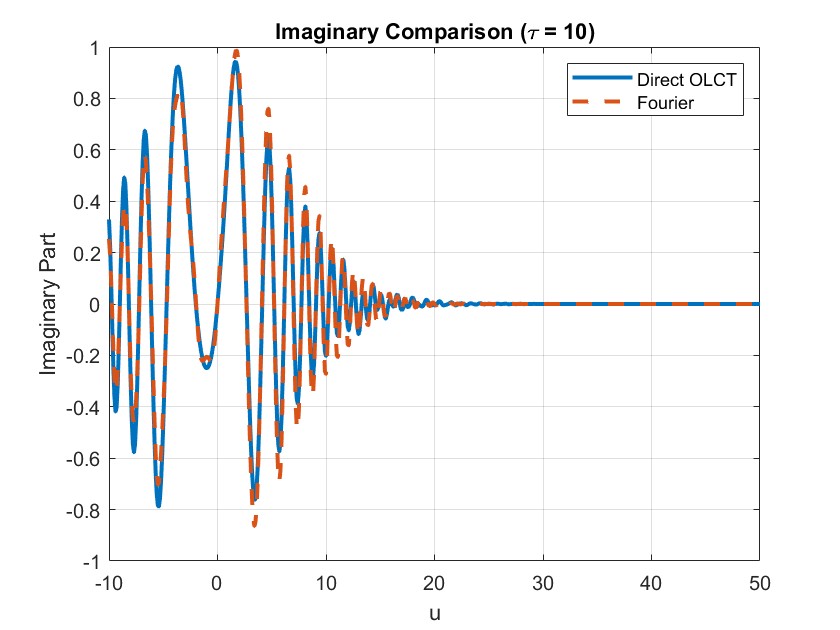} \\
(c) Real part at $\tau=10$ &
(d) Imaginary part at $\tau=10$ \\[0.3cm]

\includegraphics[width=8.5cm]{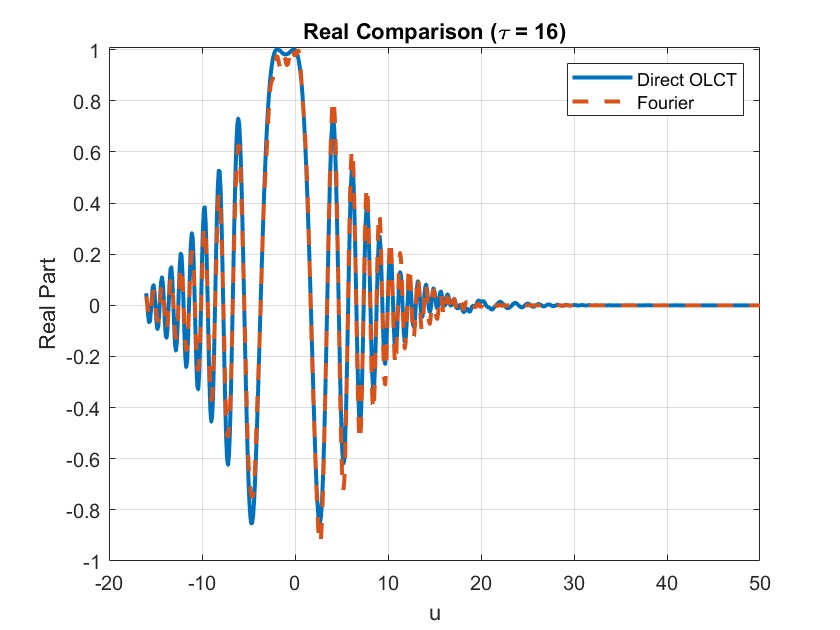} &
\includegraphics[width=8.5cm]{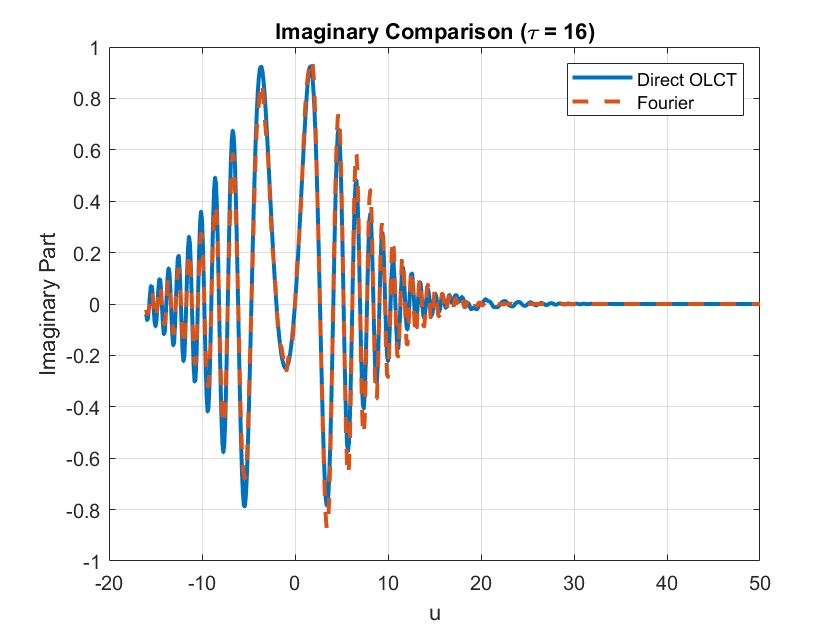} \\
(e) Real part at $\tau=16$ &
(f) Imaginary part at $\tau=16$
\end{tabular}

\caption{Comparison of the real and imaginary parts of the direct and Fourier-based solutions for $\beta<0$ at different values of the delay parameter $\tau$ for equation \eqref{olct domain solution} and \eqref{via ft solution}.}
\label{fig:beta_negative_comparison}
\end{figure}

\begin{table}[ht]
\centering
\caption{Comparison of Direct OLCT and Fourier-based solutions for different values of $\tau$.}

\renewcommand{\arraystretch}{1.4}
\begin{tabular}{|c|c|c|c|c|}
\hline
$u$ & Direct Re & Fourier Re & Direct Im & Fourier Im \\
\hline

\multicolumn{5}{|c|}{$\tau = 4$} \\
\hline
4  & 0.58787  & 0.74871  & -0.25611 & -0.30089 \\
8  & 0.08815  & 0.15582  & 0.27003  & 0.37739 \\
12 & -0.03908 & -0.05440 & -0.12608 & -0.16366 \\
16 & -0.02846 & -0.04217 & 0.02090  & 0.01359 \\
20 & -0.00087 & -0.00630 & 0.00169  & 0.00019 \\
\hline

\multicolumn{5}{|c|}{$\tau = 10$} \\
\hline
4  & 0.68279  & 0.79004  & -0.21516 & -0.26155 \\
8  & 0.12921  & 0.19036  & 0.30945  & 0.39392 \\
12 & -0.02879 & -0.06186 & -0.12622 & -0.14660 \\
16 & -0.04724 & -0.03963 & 0.01147  & 0.00873 \\
20 & -0.00916 & -0.00696 & 0.00842  & -0.00045 \\
\hline

\multicolumn{5}{|c|}{$\tau = 16$} \\
\hline
4  & 0.70759  & 0.79939  & -0.21352 & -0.24463 \\
8  & 0.15512  & 0.19465  & 0.33308  & 0.40988 \\
12 & -0.04694 & -0.06951 & -0.12340 & -0.14915 \\
16 & -0.04772 & -0.03856 & 0.00499  & 0.01087 \\
20 & -0.02157 & -0.00657 & 0.00112  & -0.00025 \\
\hline

\end{tabular}\label{table comparision of two solution}
\end{table}

\begin{figure}[h]
\centering
\begin{tabular}{cc}
\includegraphics[width=8.5cm]{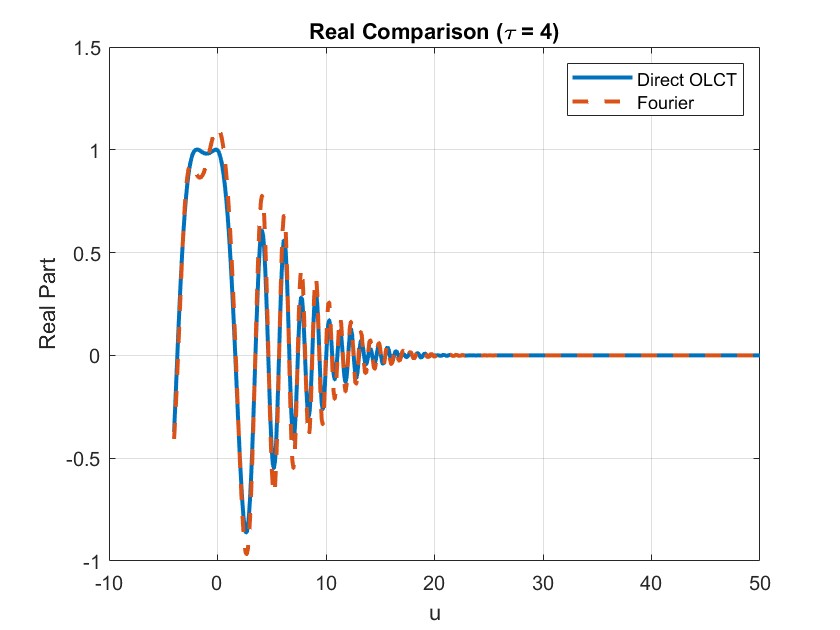} &
\includegraphics[width=8.5cm]{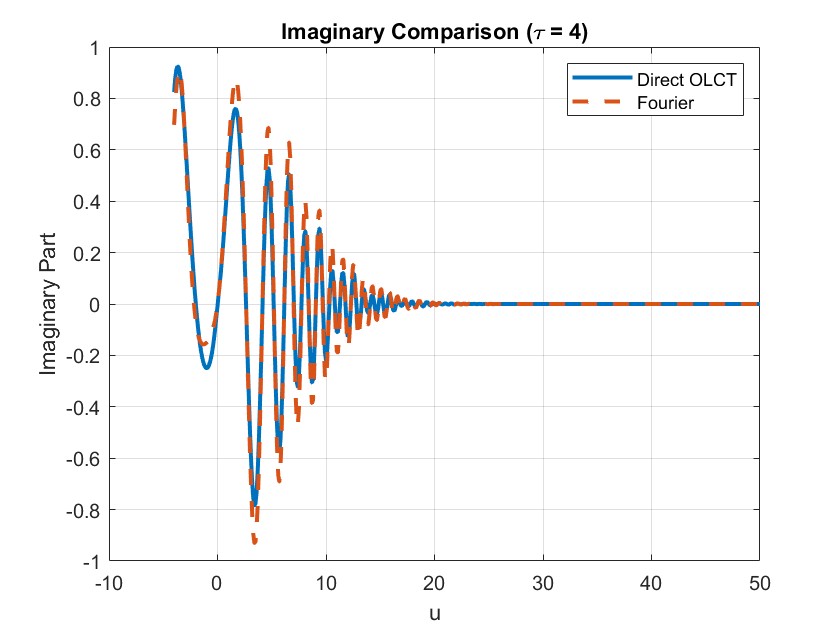} \\

(a) Real part at $\tau=4$ &
(b) Imaginary part at $\tau=4$ \\[0.3cm]

\includegraphics[width=8.5cm]{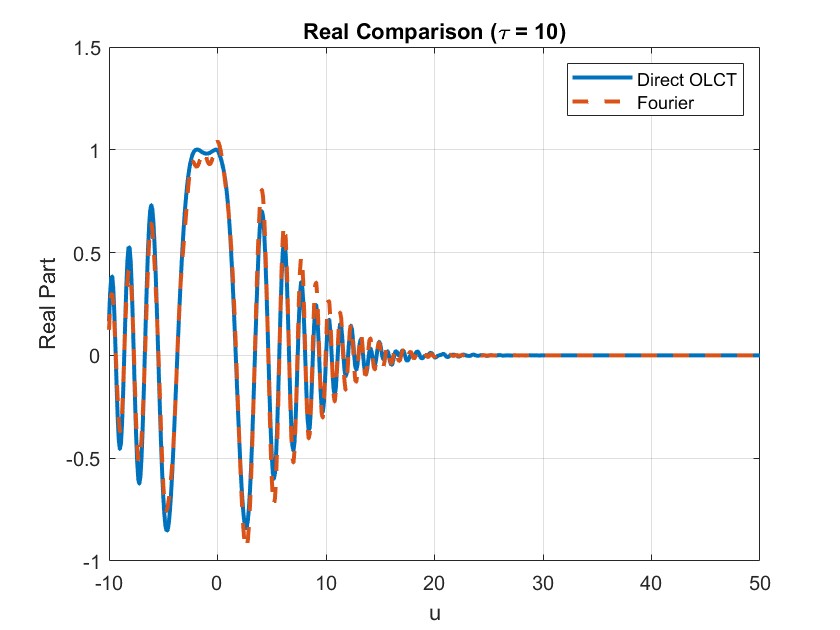} &
\includegraphics[width=8.5cm]{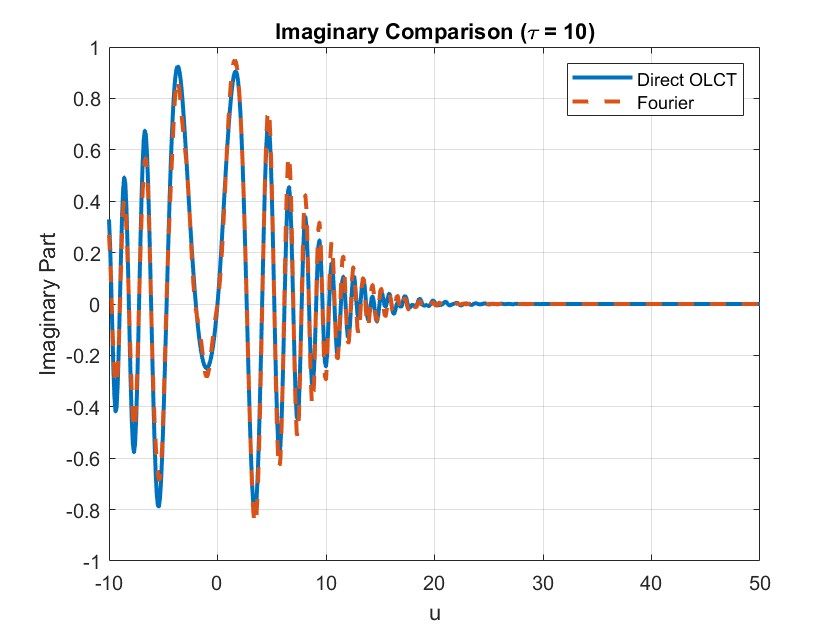} \\

(c) Real part at $\tau=10$ &
(d) Imaginary part at $\tau=10$ \\[0.3cm]

\includegraphics[width=8.5cm]{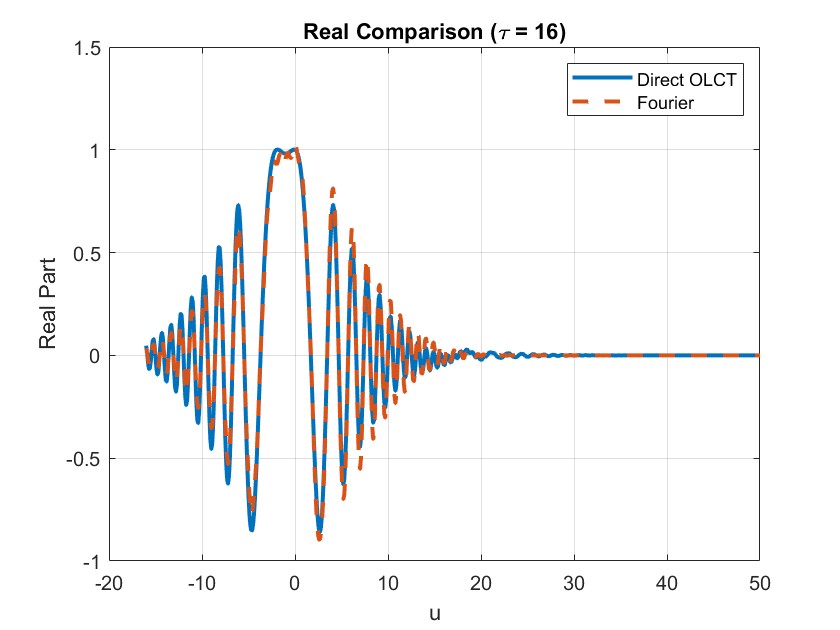} &
\includegraphics[width=8.5cm]{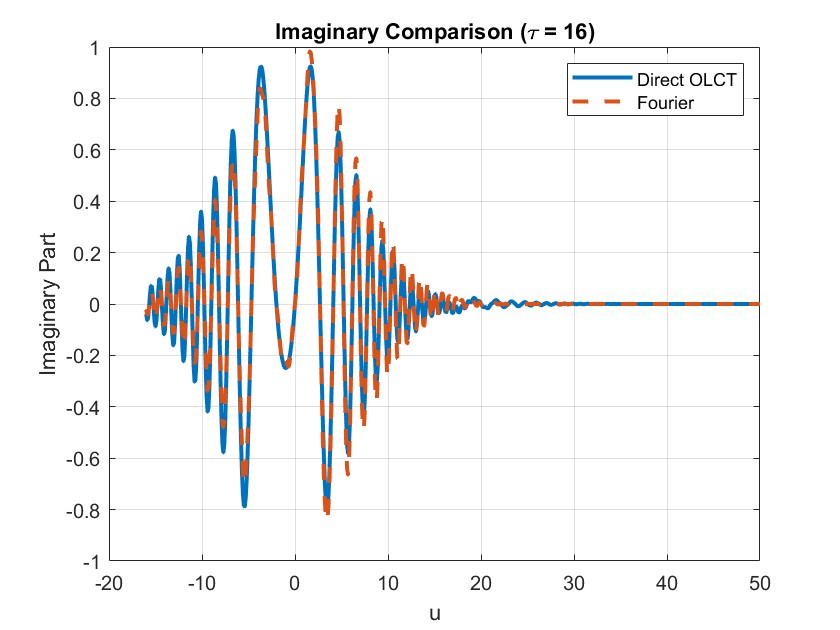} \\

(e) Real part at $\tau=16$ &
(f) Imaginary part at $\tau=16$
\end{tabular}

\caption{Comparison of the real and imaginary parts of the direct and Fourier-based solutions for $\beta>0$ at different values of the delay parameter $\tau$ for equation \eqref{olct domain solution} and \eqref{via ft solution}.}
\label{fig:beta_positive_comparison}
\end{figure}

\subsection{Observation:}Figures \ref{fig:beta_positive_comparison} and \ref{fig:beta_negative_comparison}, together with Tables \ref{table camparision of two solution negative beta} and \ref{table comparision of two solution}, compare the direct OLCT-domain solution with the Fourier-based solution for different values of the delay parameter $\tau$. It is observed that both methods produce nearly identical real and imaginary parts over the entire computational domain. The two solutions overlap closely in all cases, confirming that the proposed OLCT approach accurately reproduces the Fourier-based solution. Small differences can be observed near the oscillation peaks, where the direct OLCT solution has slightly different amplitudes. However, these differences become smaller as the solution decays. Overall, both methods show the same oscillatory pattern, phase, and decay rate for all values of $\tau$, confirming that the proposed OLCT-based approach is consistent and reliable.

\FloatBarrier
\section{Error Analysis}
The Fourier transform solution given in Equation \eqref{ohira equation} is adopted from the work of Ohira \cite{ohira ft}.
\begin{eqnarray}
X(t)
&=&
\frac{\mathcal{G}}{2\pi}
\int_{-\infty}^{\infty}
\exp\left[
-\frac{\omega^{2}}{2a}
+\frac{b}{a\tau}\cos(\omega\tau)
\right]
\cos\left(
\frac{b}{a\tau}\sin(\omega\tau)
-\omega t
\right)\,d\omega.\label{ohira equation}
\end{eqnarray}

\begin{figure}[h]

    \centering
    \begin{tabular}{cc}
    \includegraphics[width=8.5cm]{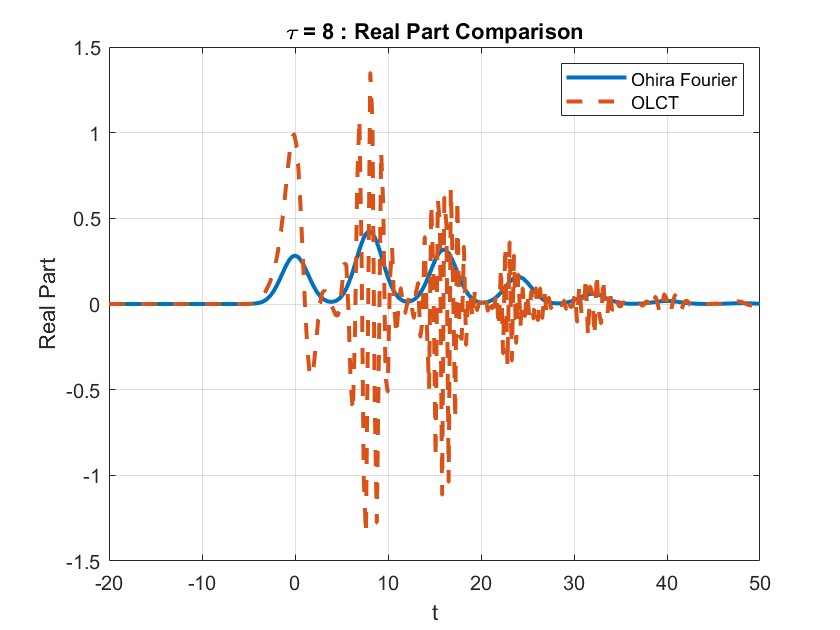}&
    \includegraphics[width=8.5cm]{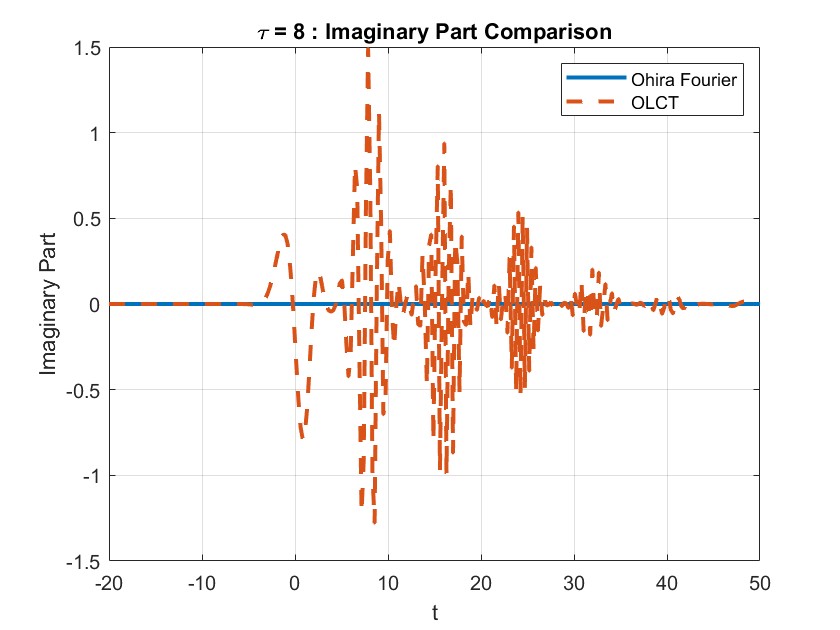}
    \label{fig:placeholder} 
    \end{tabular}
    \caption{Comapare with Fourier\eqref{ohira equation} and olct solution \ref{OLCT solution via fourier}}
\end{figure}

\begin{table}[h]
\centering
\caption{Numerical comparison between the Fourier and OLCT solutions.}
\label{tab:fourier_olct_comparison}
\renewcommand{\arraystretch}{1.2}
\begin{tabular}{|c|c|c|c|c|c|c|}
\hline
$t$ & Fourier Real & OLCT Real & Real Diff. & Fourier Imag. & OLCT Imag. & Imag. Diff.\\
\hline

$-4.5485$ & $0.0016001$ & $0.0037930$ & $0.0021930$ & $0$ & $-0.0042173$ & $0.0042173$\\
\hline
$3.4114$ & $0.0175670$ & $0.0603040$ & $0.0427370$ & $0$ & $-0.0155360$ & $0.0155360$\\
\hline
$11.1370$ & $0.0369960$ & $-0.0066718$ & $0.0436680$ & $0$ & $0.1309800$ & $0.1309800$\\
\hline
$18.8630$ & $0.0411120$ & $0.0507170$ & $0.0096057$ & $0$ & $-0.1366300$ & $0.1366300$\\
\hline
$26.5890$ & $0.0297540$ & $-0.0832270$ & $0.1129800$ & $0$ & $-0.0647930$ & $0.0647930$\\
\hline
$34.5480$ & $0.0117430$ & $0.0411350$ & $0.0293920$ & $0$ & $-0.0063813$ & $0.0063813$\\
\hline
$42.2740$ & $0.0049002$ & $0.0118300$ & $0.0069301$ & $0$ & $0.0127200$ & $0.0127200$\\

\hline
\end{tabular}
\end{table}

\subsubsection{Observation:}The Fourier solution remains purely real and exhibits a smooth Gaussian-type behavior. In contrast, the OLCT solution contains both real and imaginary components and exhibits oscillatory behavior due to the additional phase and offset parameters of the OLCT. The overall behavior of the two solutions is similar, with the main differences appearing near the oscillation peaks. The maximum absolute differences in the real and imaginary parts are 1.750497 and 1.491125, respectively. These results indicate that the OLCT formulation remains consistent with the classical Fourier-transform solution while also capturing the phase and offset effects introduced by the OLCT.

\section{Discussion and conclusion}
$1$. The analysis of delay differential equations is challenging due to the presence of delayed terms. In the proposed method, when the direct OLCT is applied to the defined delay differential equation, we obtain a Volterra-type delay integral equation. To solve this equation, we use numerical and graphical methods, since the integral equation generally does not always admit an analytical closed-form solution. In the other method, we use the relationship between the Fourier transform and the OLCT to obtain an analytical solution. Thus, using the OLCT, we can obtain both analytical and numerical solutions for delay differential equations.
\\

$2$. Several initial functions were considered to investigate the influence of the history function on the solution. The numerical approximations obtained using the proposed OLCT-based framework are consistent with the corresponding Fourier-transform solutions. The results show that changes in the initial function affect the solution behavior, while the proposed method remains applicable and accurately captures the memory effects introduced by the delay.
\\

$3$. The effectiveness of the proposed OLCT solution is examined through two comparisons. In the first comparison, the direct OLCT solution is compared with the Fourier-based OLCT solution. The numerical values and the corresponding graphs show that the two methods produce almost the same results, which confirms the correctness and consistency of the proposed approach.

 In the second comparison, the Fourier-based OLCT solution is compared with the classical Fourier-transform solution developed by Ohira. The real part of the OLCT solution matches the Fourier solution very closely. However, the Fourier solution contains only the real component, whereas the OLCT solution also provides an imaginary component because of its complex-valued kernel. This shows that the proposed OLCT method remains consistent with the classical Fourier-transform solution while providing additional information through its complex-valued representation.
\\

$4$. The numerical results and error analysis demonstrate that the proposed method provides accurate and reliable solutions for linear delay differential equations. This study serves as an initial step toward our broader objective of solving nonlinear delay differential equations involving transcendental terms. Although such problems present significantly greater mathematical challenges, the convolution and correlation properties of the Offset Linear Canonical Transform (OLCT), together with the flexibility of the resulting Volterra integral equation formulation, suggest that the proposed framework has the potential to effectively tackle these complexities. Extending the present approach to nonlinear delay differential equations with transcendental nonlinearities will be the focus of our future research.

\FloatBarrier

\bibliographystyle{amsplain}

    \end{document}